\documentclass[12pt]{article}
\usepackage{palatino}
\usepackage{a4}
\usepackage{graphicx}
\usepackage{convent}
\newcommand{\eref}[1]{(\ref{#1})}

\newcommand{\dbar}{\rule[1.25ex]{0.8ex}{0.4pt}\kern-1.2ex{\rm d}}
\usepackage{bm}
\DeclareBoldMathCommand{\bfmu}{\mu}
\DeclareBoldMathCommand{\beps}{\varepsilon}
\usepackage{bbm}
\newcommand{\tens}[1]{\mathcal{#1}}
\newcommand{\ldos}{\textsc{ldos}}

\newcommand{\bibpath}{/Users/carstenh/Biblio/Database/}
\newcommand{\bstpath}{/Users/carstenh/Biblio/Database/bst/}

\begin{document}
\sloppy

\begin{flushleft}
    {\Large
{\bf The Physics of Atom--Surface Interactions}
\\
}
\large
Carsten Henkel
\\
{\normalsize Institut f\"{u}r Physik, Universit\"at Potsdam, 
Germany
\\
{\it
Advanced School, Les Houches May 2004}}
\end{flushleft}

\noindent
In this lecture, an overview on interactions between atoms and
surfaces is given that are mediated by the electromagnetic
field. The emphasis is on dispersion (or van der Waals) forces
and transitions induced by thermal fluctuations in the near 
field of a surface. Applications with cold atoms held in
microscopic traps near surfaces, as outlined in the lecture
by J. Schmiedmayer, provide the experimental background for
this theory lecture.

We do not cover the chemical physics aspects that are relevant
at distances comparable to the atomic scale. Phenomena like
adsorption and diffusion on surfaces, or atomic beam diffraction
from crystalline surfaces, are left to the lectures by J. Frenken
and H. Lezec.

\section*{Outline}

\paragraph{First part: forces}

\begin{itemize}
\item
Theory of the Van der Waals interaction. 
Electric dipole coupling, force from electromagnetic field
fluctuations.
\item
Characterization of the field near a surface. Cross spectral density,
fluctuation-dissipation theorem, example: planar surface.
\item
Discussion of the Van der Waals--Casimir force. Asymptotic
calculation of the near field spectral density. 
\end{itemize}

\paragraph{Second part: transitions}

\begin{itemize}
\item
Quantum states relevant for atoms

\item
Fermi's Golden Rule, connection to the cross spectral density

\item
Rates for different processes, scaling laws.

\end{itemize}

\section{Van der Waals Forces}

\subsection{Conventional viewpoint}

The Van der Waals force derives from the energy shift 
an atomic ground state experiences in front of a surface.
This shift is computed in second order
perturbation theory, coupling the atom to the mode continuum of the
electromagnetic field. The following section ({\small in small characters
like these}) is taken from the lecture 
notes ``Theoretical Quantum Optics I'', held at Universit\"{a}t 
Potsdam (Germany) winter semester 2003/04. The complete notes are on line at 
{\tt www.quantum.physik.uni-potsdam.de}.

\begin{small}

\subsubsection{Single-mode vacuum shift}

Let us start with a single mode of the electromagnetic field.
We have seen previously that the ground state
$|g;0\rangle$ of the atom+field system
is not affected by the interaction. This is actually 
only true in the rotating wave approximation. When we include the 
nonresonant terms%
\footnote{%
$g$ is a coupling constant with dimensions frequency, $a$ and 
$a^\dag$ are the annihilation and creation operators for a photon in 
the field mode, $\sigma$ and $\sigma^\dag$ are the corresponding
atomic ladder operators.},
\[
H_{\rm nr} = \hbar g ( \sigma_{+} a^\dag + \sigma_{-} a )
\]
they do affect the ground state, since
\[
H_{\rm nr} | g; 0 \rangle = \hbar g | e; 1 \rangle
.
\]
In second order perturbation theory, we therefore get the 
following energy shift
\begin{eqnarray}
\delta E_{g0} &=& 
- \frac{ \left| \langle e;1 | H_{\rm nr} | g;0 
\rangle \right|^2}{ E_{e1} - E_{g0} }
=
- \frac{ \hbar^2 g^2 }{ \hbar ( \omega_{A} + \omega ) }
\label{eq:Lamb-single-mode}
\end{eqnarray}
This shift is very small compared to those due to the resonant 
interaction because here the ``detuning'' $\omega_{A} + \omega$
is large (the transition can only happen ``virtually'' because it 
violates energy conservation).

Nevertheless, we learn that the atom-field interaction does displace 
the absolute ground state of the system, even for a single mode.
The Lamb shift is the generalisation of this result for the full, 
multi-mode electromagnetic field.

\subsubsection{Lamb shift, a first glimpse}

The calculation of the Lamb shift is a basic example of the different 
infinities that occur in quantum electrodynamics. We only give a 
first idea of the calculation, without going into the many technical 
details (renormalisation, subtraction of classical energy shifts like 
Coulomb and polarisation self-interaction etc.).

For the multi-mode field, any one-photon state $| e; 1_{{\bf k}\mu} 
\rangle$ is coupled via the nonresonant interaction to the ground 
state $|g; 0\rangle$, giving a coupling matrix element $\hbar g_{{\bf 
k}\mu}$. Summing over all these modes, we find the energy shift
\begin{eqnarray}
    \delta E_{g0} &=& - \frac{1}{\hbar}
    \sum_{{\bf k}\mu} \frac{ \hbar^2 |g_{{\bf k}\mu}|^2 }{ \omega_{A} + 
    \omega_{k} }
    \nonumber\\
    &=&
    - \frac{1}{\hbar}
    \sum_{{\bf k}\mu} \frac{
    E_{k}^2 
    | {\bf d} \cdot 
    \mbox{\boldmath$\varepsilon$}_{{\bf k}\mu}|^2
    | f_{{\bf k}\mu}( {\bf x} ) |^2 }{ 
    \omega_{A} + k c }
    \label{eq:Lamb-shift}\\
    &=&
    - \frac{1}{\hbar}
    \frac{ V }{ (2\pi)^3 }
    \int\!{\rm d}^3k \,
    \frac{ \hbar k c }{ 2 \varepsilon_{0} V ( \omega_A + k c ) }
    \left(
    | {\bf d} |^2
    - 
    | {\bf d}\cdot \hat{\bf k} |^2 
    \right)
    | f_{{\bf k}}( {\bf x} ) |^2
\nonumber
\end{eqnarray}
We have introduced the mode functions $f_{{\bf k}\mu}( {\bf x} )$ 
(equal to $\exp( {\rm i} {\bf k} \cdot {\bf x} )$ for a plane wave 
expansion) and, assuming that $| f_{{\bf k}}( {\bf x} ) |^2$ is
independent of $\mu$, 
 performed the summation over the polarization unit 
vectors (as already seen a few times in the exercises). Good luck,
the volume $V$ cancels with the factor $V^{-1/2}$ from the `field per
photon' $E_{k}$. The 
integration over the angles of ${\bf k}$ gives:
\begin{equation}
\int\!\sin\theta \,{\rm d}\theta\,{\rm d}\varphi \,
    \left(
    | {\bf d} |^2
    - 
    | {\bf d}\cdot \hat{\bf k} |^2 
    \right)
=
\frac{8\pi}3 | {\bf d} |^2
.
\label{eq:angular-int}
\end{equation}
We are finally left with the radial integral
\begin{equation}
    \delta E_{g0} = - \frac{ 8 \pi }{6 (2\pi)^3 }
    \frac{  
    | {\bf d} |^2 }{ \varepsilon_{0} }
    \int_{0}^\Lambda\!
    \frac{ k^3 \,{\rm d}k }{ \omega_{A}/c + k }
\qquad (\mbox{wrong})
    \end{equation}
This integral is obviously divergent in the limit $k \to \infty$, 
which is called an ``ultraviolet catastrophe''. The integral 
is only finite if 
we introduce a cutoff wavenumber $\Lambda$. What order of magnitude can we 
reasonably give to $\Lambda$? Remember that our whole theory is 
based on the ``long-wavelength approximation'' (photon wavelength 
$2\pi/k$ much larger than atom size $\sim a_{0}$). For wave vectors
$k \gg 1/a_{0}$ we therefore cannot trust our interaction Hamiltonian 
any more. Choosing the cutoff $\Lambda = 1/a_{0}$, we get a shift of 
the order of $d^2 \Lambda^3 / \varepsilon_{0} = d^2 / \varepsilon 
a_{0}^3 \sim e^2 / \varepsilon_{0} a_{0}$. This is a \emph{completely wrong}
result because it is of the order of the atomic binding energy! 

To get the right result, one has to take into account self-energy terms that 
we did not explicitly write down in the Hamiltonian. These are infinite, too, 
but if they are evaluated with a similar cutoff and subtracted, one gets 
something which does not too badly diverge with the cutoff $\Lambda$:
\begin{equation}
\delta E_{g0} \approx
\frac{ 1 }{ 3\pi^2 }
\frac{ | {\bf d}|^2 \omega_A^3 }{ \varepsilon_0 c^3 }
\ln\!\left( \frac{ \Lambda }{ \omega_A / c }
\right)
\qquad (\mbox{correct})
\end{equation}
Here, the correct value of the cutoff is given by \ldots
the Compton wavevector of the electron, $\Lambda = m c / \hbar $ 
(up to this wavevector 
is our nonrelativistic theory valid).

It is beyond the scope of this lecture to give a more detailed
account of the Lamb shift calculation.


\subsubsection{Van der Waals and Casimir-Polder forces}

The previous calculation, as sick as it is, does give a result for the
\emph{modification} of the Lamb shift due to boundary conditions imposed
on the electromagnetic field by macroscopic bodies. Let us focus on the 
simple case that our two-level atom is placed at a distance $z$ from a
perfectly reflecting mirror. All we have to change in the calculation
are the mode functions, and we shall use
\[
f_{{\bf k}\mu}( {\bf x} ) = \sqrt{2} 
\sin(k_z z) \exp[ {\rm i} ( k_x x + k_y y ) ]
.
\]
(We are actually cheating with the polarisation vectors to simplify things.)
The factor $\sqrt{2}$ comes from the fact that we use a unitary 
transformation%
\footnote{%
Alternative argument: normalise the volume integral 
of the squared mode function to $V$.}
of the plane waves ${\rm e}^{ \pm {\rm i} k_z z}$. We have to change
the angular integration~(\ref{eq:angular-int}) because now 
the $z$-axis plays a
privileged role. We can expand the dipole moment in cartesian
components and get 
($\theta$ is now the angle between ${\bf k}$ and the $z$-axis):
\begin{eqnarray*}
\int\!{\rm d}\varphi \, 
| {\bf d}\cdot \hat{\bf k} |^2 
&=&
\int\!{\rm d}\varphi \, 
\Big| d_x \sin\theta \cos\varphi + 
d_y \sin\theta \sin\varphi + 
d_z \cos\theta \Big|^2
\\
&=&
2\pi \left(
{\textstyle\frac12} d_x^2 \sin^2\theta 
+
{\textstyle\frac12} d_y^2 \sin^2\theta 
+
d_z^2 \cos^2\theta 
\right)
\end{eqnarray*}
Note that the integrals over the mixed terms involve $\cos\varphi$,
$\sin\varphi$, or $\cos\varphi \sin\varphi$ whose average over one
period vanishes. In terms of the components 
$d_\Vert = d_z$ and $d_\perp$ of the dipole parallel and perpendicular 
to the $z$-axis, we thus get
\begin{eqnarray}
\delta E_{g0} &=&
- 
\frac{ 1 }{ 2 ( 2\pi )^2 
\varepsilon_0 } 
\int\!\frac{
k^3 {\rm d}k \,\sin\theta {\rm d}\theta}{
\omega_a / c + k }
\nonumber
\\ && \quad {}\times
\left(
d_\Vert^2 \sin^2\theta + d_\perp^2 ( 1 - {\textstyle\frac12}\sin^2\theta )
\right)
\left( 1 - \cos( 2 k z \cos\theta ) \right)
\end{eqnarray}
The `1' in the last bracket is independent of the distance $z$ to the mirror 
--- it will give the Lamb shift for an atom in free space. The second term
with the $\cos$ gives the modification of the Lamb shift we are interested
in. 
The $\theta$ integral
gives (make the substitution $ u = \cos\theta$ and differentiate with 
respect to $2kz$ to get the integrals with the additional factor 
$\cos^2\theta$)
\begin{eqnarray*}
\int_0^\pi\!\sin\theta\,{\rm d}\theta \,
\cos( 2 k z \cos\theta )
& = &
\frac{ 2 \sin 2 k z }{ 2 k z }
\\
\int_0^\pi\!\sin\theta\,{\rm d}\theta \,
\cos^2\theta 
\cos( 2 k z \cos\theta )
& = &
\frac{ 2 \sin 2 k z }{ 2 k z } +
\frac{ 4 \cos 2 k z }{ (2 k z)^2 } -
\frac{ 4 \sin 2 k z }{ (2 k z)^3 }
\\
\int_0^\pi\!\sin\theta\,{\rm d}\theta \,
\sin^2\theta 
\cos( 2 k z \cos\theta )
& = &
- \frac{ 4 \cos 2 k z }{ (2 k z)^2 } +
\frac{ 4 \sin 2 k z }{ (2 k z)^3 }
\end{eqnarray*}
Apparently, we get an expression with different powers of $1/z$. 

Let us focus first on the \textbf{short-distance limit} and consider
only the ``most diverging'' term $\sim 1/z^3$.  It involves the
integral
\[
I_3(z) = 
\int_0^\infty\!\frac{ {\rm d}k \, \sin 2 k z }{ k_A + k }
\]
where we have put $k_A = \omega_A / c$. We use a trick in the complex
plane to perform the integration: write $\sin 2 k z$ as the imaginary
part of ${\rm e}^{ 2 {\rm i} k z }$ and deform the integration path in
the complex plane from the positive real axis to the positive imaginary
axis. This works because the exponential ${\rm e}^{ 2 {\rm i} k z }$
vanishes at infinity, and the integrand has no pole in the upper left
quadrant. Using the integration variable $k = {\rm i}\kappa$
on the imaginary axis, we find
\begin{eqnarray}
I_3( z ) & = & {\rm Im}\,
\int_0^\infty\!\frac{ {\rm d}k \, {\rm e}^{ 2 {\rm i} k z }  }{ k_A + k }
=
{\rm Im}\,
{\rm i}\int_0^\infty\!\frac{ {\rm d}\kappa \, {\rm e}^{ -2 \kappa z  } }
{ k_A + {\rm i}\kappa }
\nonumber
\\
&=&
k_A \int_0^\infty\!\frac{ {\rm d}\kappa \, {\rm e}^{ -2 \kappa z  } }
{ k_A^2 + \kappa^2 }
\label{eq:two-scale-integral}
\end{eqnarray}
We approximate this integral by noticing that it is the product of
two decaying functions. 
%
%
The exponential decays on a scale given by
$1/z$, while the ``Lorentzian'' $1/(k_A^2 + \kappa^2)$ decays on the 
scale $k_A$. Now look at which scale is larger. The short distance
regime is defined by the limit $z \ll 1/k_A$: the distance is much 
smaller than the transition wavelength. Then the exponential decays much 
more slowly, and we can replace it by unity. This gives
\[
k_A z \ll 1: \quad
I_3(z) = 
k_A \int_0^\infty\!\frac{ {\rm d}\kappa  }
{ k_A^2 + \kappa^2 }
= \arctan(\kappa/k_A)\Big|_0^\infty = \frac{\pi}{2}
\]
The other inverse powers can be handled in a similar way and give
sub-dominant contributions. We thus find the so-called van der Waals 
energy
\begin{equation}
E_{\rm vdW} = -
\frac{ 1 }{ 4 \pi\varepsilon_0  }
\frac{  
d_\Vert^2 - {\textstyle\frac12} d_\perp^2 
}{ 8 z^3 } 
\qquad (z \ll \lambda_A, \mbox{ wrong} ).
\end{equation}
This is quite close to the correct result.
The correct calculation with the proper hand\-ling of the polarisation
vectors gives
\begin{equation}
E_{\rm vdW} = 
-
\frac{ 1 }{ 4 \pi\varepsilon_0  }
\frac{  
d_\Vert^2 + {\textstyle\frac12} d_\perp^2 
}{ 8 z^3 }
\qquad (z \ll \lambda_A,  \mbox{ correct}).
\end{equation}
which differs only in the sign of the $d_\perp^2$ term.
For realistic atoms, you get contributions from all excited states that
are connected to the ground state (and from all their Zeeman sublevels).

In the \textbf{long-distance limit} $k_A z \gg 1$ (distance larger
than transition wavelength), we can also evaluate the
integral~(\ref{eq:two-scale-integral}).  Now the exponential has the
fastest decay, and we can replace the lorentzian by a constant:
\[
k_A z \gg 1: \quad
I_3(z) = 
\frac{ 1 }{ k_A } \int_0^\infty\!{\rm d}\kappa \,
{\rm e}^{ - 2 \kappa z }
= \frac{1 }{2 k_A z}
,
\]
and hence a result much smaller than in the short 
distance limit (because now $k_{A} z \gg 1$). This occurs
in fact for all terms we got. The final result is the famous Casimir-Polder
energy:
\[
E_{\rm CP} = 
- \frac{ 1 }{ 4\pi\varepsilon_0 }
\frac{ {\bf d}^2 }{ 4 \pi k_A z^4 } 
\qquad (z \gg \lambda_A ).
\]
where ${\bf d}^2 =  d_\Vert^2 + d_\perp^2$ (no distinction between
the dipole orientations).  

\end{small}

\subsubsection{Summary}

An atom in its ground state close to a planar surface feels 
an attractive potential
that scales like $1/z^3$ (van der Waals potential).
This applies to distances much shorter than the
wavelength $\lambda_A$ of the strongest electric dipole transition 
starting from the ground state. The coefficient depends on the
squared dipole moment ${\bf d}^2$ of the atom, 
more precisely its expectation
value in the ground state. At large distances $z \gg \lambda_A$,
a power law $1/z^4$ is found (Casimir-Polder potential). The
coefficient depends on ${\bf d}^2 / \omega_A$, and this quantity
is proportional to the atom's static polarizability $\alpha(0)$.
More details on this calculation can be found, for example, in
the textbook by Craig and Thirunamachandran~\cite{CraigThiruna}
and the 1990 Les Houches lecture by Haroche~\cite{Haroche92}.

For later reference, we quote here the result for the polarizability of
the atomic ground state. This formula can be found by 
evaluating to lowest order in an applied electric field (frequency
$\omega$) the average dipole induced moment.
\begin{equation}
    \alpha_{ij}( \omega ) = \frac{2}{\hbar}
    \sum_{e} \frac{ \omega_{eg} 
    \langle g | d_{i} | e \rangle \langle e | d_{j} | g \rangle }{
    \omega_{eg}^2 - \omega^2 - 0 {\rm i}\,\omega }
    \label{eq:alpha-ground-state}
\end{equation}
Here, the Bohr frequency between the ground state $g$ and some 
excited state $e$ is $\omega_{eg} = (E_{e} - E_{g}) / \hbar$. The 
infinitesimal imaginary term in the denominator ensures causality: as 
a physical response function, the polarizability $\alpha_{ij}( 
\omega )$ can only have poles in the lower half of the complex 
$\omega$-plane. Note that a number of different conventions for the 
units of the polarizability occur in the literature. In our notation, 
$\alpha_{ij}( \omega ) / \varepsilon_{0}$ has the units of volume, and
the average induced dipole moment is given by
$\langle {d}_{i} \rangle {\rm e}^{ - {\rm i} \omega t} + \mbox{c.c.}
= \alpha_{ij}( \omega ) {E}_{j}( {\bf r} ) \,
{\rm e}^{ - {\rm i} \omega t} + \mbox{c.c.}$
(Here and in the following, we use Einstein's
summation convention and drop the sum $\sum_{j}$.)

For the static polarizability, we get
\[
\alpha_{ij}( 0 ) = 
\sum_{e}\frac{ 2
    \langle g | d_{i} | e \rangle \langle e | d_{j} | g \rangle }{
    \hbar \omega_{eg} }
.
\]
If the ground state has spherical symmetry, this tensor is actually 
proportional to $\delta_{ij}$. (Note that in the hydrogen atom, this 
requires the summation over the three $p_{x}, p_{y}, p_{z}$ excited 
states.)

\subsection{Role of field fluctuations}

We now present an alternative viewpoint on the van der Waals force.
Its advantage is that the role of field fluctuations is made
explicit. Both vacuum and thermal fluctuations contribute at 
finite temperature. We develop a formalism that treats both
in a unified way. At zero temperature, the standard van der Waals
force is recovered. A thermal correction appears in the near field
of a `hot' surface. 

The following section is adapted from the paper ``Radiation forces on small 
particles in thermal near fields'' by C.\ Henkel, 
K.\ Joulain, J.-P.\ Mulet, and J.-J.\ Greffet [J.\ Opt.\ A: Pure 
Appl.\ Opt.\ {\bf 4} (2002) S109].

\begin{small}

\subsubsection{Radiation force}

We consider a model system made from a dielectric half-space
and a small (sub-wavelength) ``test particle'' in vacuum above. 
Both objects
are possibly at finite temperature and radiate electromagnetic
fields whose sources are thermal current fluctuations. Using
the fluctuation electrodynamics framework of Lifshitz 
\cite{Lifshitz56}, the spectral density of these
currents is related to the imaginary part of the dielectric function
(for the half-space) or of the polarizability (for the particle).

We are interested in the force exerted by the radiation on the
test particle. This force is actually both a time-average over 
the rapid fluctuations of the thermal field and an average over
a statistical ensemble for the field. For particles with
a subwavelength size, the force is given by \cite{Gordon80}
\begin{equation}
    {\bf F}( {\bf r} ) = 
    \sum_{i=x,y,z}\langle d_{i}(t) \nabla E_{i}( {\bf r}, t) \rangle
    \label{eq:def-force}
\end{equation}
where ${\bf d}(t)$ is the dipole moment of the particle
and ${\bf E}({\bf r},t)$ the total electric field at the 
particle's position. The latter contains the fields
generated by the thermal substrate and by the particle's 
dipole moment, as well as the blackbody field in the vacuum 
half-space. The
expression \eref{eq:def-force} 
combines both the Coulomb force (corresponding to the
potential energy $- {\bf d} \cdot {\bf E}$) and the Lorentz force
(involving the magnetic field), see \cite{Gordon80}. 
As a simple derivation, we note that the force operator can be defined 
in terms the 
Heisenberg equation for the momentum operator
\begin{equation}
    \hat {\bf F} = \dot {\bf p} =
    \frac{ {\rm i} }{ \hbar } \left[ H, {\bf p} \right]
     = - \nabla H_{\rm int} 
     \label{eq:derive-force}
 \end{equation}
where we have used the fact that the atomic Hamiltonian is independent 
of the atomic center-of-mass position. Using the electric dipole interaction,
one gets~(\ref{eq:def-force}). 

In our case, both the dipole moment and the electric field
are fluctuating quantities. The force may therefore
be written as a sum of two terms
\begin{equation}
    {\bf F}( {\bf r} ) = 
    \langle d^{(\rm ind)}_{i}(t) 
    \nabla E^{(\rm fl)}_{i}( {\bf r}, t) \rangle
    +
    \langle d^{(\rm fl)}_{i}(t) 
    \nabla E^{(\rm ind)}_{i}( {\bf r}, t) \rangle
    \label{eq:two-terms}
\end{equation}
where the first describes the (spontaneous and thermal) fluctuations 
of the field that correlate with the corresponding induced dipole, 
while the second involves dipole fluctuations and the field they
induce.
There is no term involving fluctuations of both the dipole and the
field: these are not correlated, since they originate from different
physical systems. More precisely, 
to lowest order in perturbation theory,
we can take the fluctuations in 
\eref{eq:two-terms} to be those of the non-coupled atom+field system,
these being obviously uncorrelated.

\subsubsection{Induced fluctuations}

The dipole induced by the field fluctuations is given by the 
particle's polarizability:
\begin{equation}
    {\bf d}^{(\rm ind)}( \omega ) = 
    \alpha( \omega ) 
    {\bf E}( {\bf r}; \omega )
    \label{eq:induced-dipole}
\end{equation}
where we switched to frequency Fourier transforms and 
assumed an isotropic polarizability. Note also that 
${\bf E}( \omega; {\bf r} )$ is the total electric field at the
dipole's location. If we work at lowest order in the polarizability,
we can identify this field with the fluctuating field 
${\bf E}^{(\rm fl)}( {\bf r}; \omega )$ and ignore the scattering of 
this field by the particle. Otherwise, we could work with a 
``dressed'' polarizability. Note however that the fluctuating 
field is not simply a collection of plane waves, but
takes into account their scattering from the substrate.

The field induced by the dipole fluctuations is given by the 
electromagnetic Green tensor:
\begin{equation}
    {\bf E}^{(\rm ind)}( {\bf x}; \omega ) = 
    \tens{G}( {\bf x}, {\bf r}; \omega ) \cdot
    {\bf d}( \omega )
    ,
    \label{eq:induced-field}
\end{equation}
where the variables ${\bf x}$ and ${\bf r}$ represent the observation
and source point, respectively. To lowest order, we can identify 
the total dipole in~\eref{eq:induced-field} with its fluctuating component,
as we did before for the field.

Finally, we express the operator product in terms of frequency Fourier 
transforms as
\begin{equation}
    \langle A(t) B(t) \rangle = 
    \int\!\frac{{\rm d}\omega}{2\pi} \frac{{\rm d}\omega'}{2\pi}
    \,{\rm e}^{ {\rm i} (\omega - \omega') t}
    \langle A^\dag(\omega) B(\omega') \rangle 
    \label{eq:intro-domega}
\end{equation}
where the frequency integrals run over both positive and negative 
frequencies. Without loss of generality, we have written the 
frequency integral for $B(t)$ in terms of the conjugate operator
$B^\dag[\omega]$. This is possible because in~\eref{eq:two-terms},
both $A(t)$ and $B(t)$ are real (hermitian) operators.

\subsubsection{Fluctuation spectra}

\paragraph{Dipole.}

The dipole fluctuations of the particle are characterized by a
spectral density given by the fluctuation--dissipation theorem. 
This assumes the particle to be at thermal equilibrium (temperature 
$T_{2} / k_{B}$) and gives \cite{Agarwal75a,Sipe84}
\begin{equation}
    \langle d_{i}^{\dag}(\omega) d_{j}(\omega') \rangle
    = 2\pi \delta(\omega - \omega') \delta_{ij}
    \frac{ 2\hbar 
                  }
    { {\rm e}^{\hbar \omega / T_{2}} - 1 }
    {\rm Im}\,\alpha(\omega)
    \label{eq:FD-dipole}
\end{equation}
Note that this expression applies to both positive and negative 
frequencies: for $\omega > 0$, we have $({\rm e}^{\hbar \omega / 
T_{2}} - 1)^{-1} = \bar n( \omega / T_{2} )$, 
the average excitation 
number for Bose-Einstein statistics (vanishing at $T = 0$). At
negative frequencies, $\omega < 0$, we have 
$({\rm e}^{\hbar \omega / T_{2}} - 1)^{-1} = - ( \bar n(|\omega| / 
T_{2}) + 1)$,
and the minus sign is compensated by the symmetry relation
${\rm Im}\,\alpha_{ij}(\omega) = -{\rm Im}\,\alpha_{ij}[-\omega]$
so that the total expression is positive again (as it should). 
In particular, at $T_{2}=0$, the dipole ``fluctuates
only at negative frequencies'' which reflects the existence of
the ground state (fluctuations can only connect to higher-lying
states). We identify these fluctuations with spontaneous 
fluctuations. Thermal fluctuations are proportional to $\bar n$
and are the same for positive and negative frequencies (as they
should for classical noise spectra). 

\paragraph{Field.}

For the field fluctuations, the fluctuation-dissipation theorem
reads (field at temperature $T_1$) \cite{Agarwal75a,Sipe84}
\begin{equation}
    \langle E_{i}^{\dag}({\bf r}; \omega) E_{j}({\bf r}'; \omega')
    \rangle = 
    2\pi\delta(\omega - \omega')
    \frac{ 2\hbar }
    { {\rm e}^{\hbar \omega / T_1} - 1 }
    {\rm Im}\,G_{ij}({\bf r}, {\bf r}'; \omega)
    \label{eq:FD-field}
\end{equation}
where the imaginary part of the Green tensor appears. This result
is valid at equilibrium (field and source at the same temperature).
For a generalization to non-equilibrium, see below.

\subsubsection{Resulting force}

Collecting the different contributions to the force, we show that 
the ``vacuum'' fluctuations of both particle and substrate conspire 
to give the standard van der Waals--Casimir--Polder--London force. 
The 
thermal fluctuations give a correction that we evaluate in the limit 
of a particle at zero temperature.


Starting from the general expression~\eref{eq:two-terms}, 
the force due to dipole fluctuations is
(using the notation $\dbar\omega = {\rm d}\omega / 2\pi$)
\begin{equation}
    \langle d^{(\rm fl)}_{i}(t) 
    \nabla E^{(\rm ind)}_{i}( {\bf r}, t) \rangle    
    =
    2 \hbar 
    \int\! \frac{ \dbar\omega }{
    {\rm e}^{\hbar \omega / T_{2}} - 1 }
    \left( {\rm Im}\,\alpha(\omega) \right)
    \nabla_{1} G_{ii}({\bf r}, {\bf r}; \omega)
    \label{eq:force-dip}
\end{equation}
where $\nabla_{1}$ is shorthand for the gradient with respect to the
first position variable of the Green tensor.  

The force due to field fluctuations is
\begin{equation}
    \langle d^{(\rm ind)}_{i}(t) 
    \nabla E^{(\rm fl)}_{i}( {\bf r}, t) \rangle_{{\rm vac}}
    =
    2 \hbar 
    \int\! \dbar\omega \,
    \int\! \frac{ \dbar\omega }{
    {\rm e}^{\hbar \omega / T_{1}} - 1 }
    \alpha^*(\omega)
    \left( {\rm Im}\,
    \nabla_{2} G_{ii}({\bf r}, {\bf r}; \omega)
    \right) 
    \label{eq:force-field-T1=0}
\end{equation}
We now argue that the Green tensor satisfies the symmetry relation
\begin{equation}
\nabla_{1}G_{ii}({\bf r}, {\bf r}; \omega) = 
\nabla_{2}G_{ii}({\bf r}, {\bf r}; \omega)
.
\label{eq:symmetry-Green}
\end{equation}
Very generally, the reciprocity theorem implies that 
the Green tensor is symmetric under combined exchange 
of the position arguments and vector indices \cite{Agarwal75a}. 
For the particular case 
of the half-space, we have checked this explicitly. In this case,
we also find that only 
the part involving the reflection from the interface remains
under the derivative. This is because the (imaginary part of)
the free space Green tensor depends only on 
$({\bf r} - {\bf r}')^2$, due to isotropy, and
has vanishing first derivative at ${\bf r} = {\bf r}'$. 

Using these results and specializing to the equilibrium case
$T_1 = T_2 = T$, we can combine~\eref{eq:force-dip},
\eref{eq:force-field-T1=0} to give the van der Waals force
\begin{equation}
    {\bf F}_{\rm vdW}({\bf r}) = 
    2 \hbar 
    \,{\rm Im} 
    \int\! \frac{ \dbar\omega }{
    {\rm e}^{\hbar \omega / T} - 1 }
    \alpha(\omega) \,
    \nabla_{1} G_{ii}({\bf r}, {\bf r}; \omega)
.
    \label{eq:recover-vdW-finite-T}
\end{equation}
You will find frequently in the literature an expression where
the $\omega$-integral is transformed in the complex $\omega$-plane. 
One
closes the integration path with a semicircle in the upper half plane.
This circle does not contribute to the integral because $\alpha( \omega )$
and $G_{ij}( \omega )$, being physical
response functions, decay to zero for 
$|\omega| \to \infty$. In addition, they have no singularities in the
upper half plane. The integral along the closed contour is thus
determined by the poles of the Bose-Einstein factor at $\omega 
= {\rm i}\xi_n \equiv 2\pi {\rm i} n T/\hbar, \: n = 0, 1, 2 \ldots$ 
(`Matsubara frequencies'). The pole $n = 0$ contributes with only
one half of its residue because it lies on the original contour. 
One thus gets:
\begin{equation}
    {\bf F}_{\rm vdW}({\bf r}) = 
    2 T 
{\sum_{n}}'
    \alpha({\rm i}\xi_n) \,
    \nabla_{1} G_{ii}({\bf r}, {\bf r}; {\rm i}\xi_n)
.
    \label{eq:recover-vdW-finite-T-Matsubara-sum}
\end{equation}
(The prime reminds that $n=0$ is counted with a factor $\frac12$.)
We have dropped the `${\rm Im}$' because the response functions are real 
at purely imaginary frequencies.

\paragraph{Zero temperature.}

In the limit of a
particle at zero temperature, the Bose-Einstein distribution
becomes a step function,
\begin{equation}
\lim_{T_1 \to 0}
\frac{ 1 }{ {\rm e}^{ \hbar\omega / T_1 } - 1 }
= H(-\omega),
\label{eq:Heaviside}
\end{equation}
showing that 
the integral runs only over negative frequencies.  We thus find
\begin{equation}
    {\bf F}_{\rm vac}({\bf r}) = 
    2 \hbar 
    \,{\rm Im} 
    \int_{0}^{\infty}\! \dbar\omega \,
    \alpha(\omega) \,
    \nabla_{1} G_{ii}({\bf r}, {\bf r}; \omega)
    ,
    \label{eq:recover-vdW}
\end{equation}
where we have used the symmetry properties of $\alpha$ and $G_{ij}$.
This result resembles up to a factor of $2$ and a $-\nabla_{1}$
the expression for the van der Waals--Casimir--Polder--London energy
given in~\cite{Sipe84}.
The gradient occurs because we calculate the force, of course.
To understand the factor 2, we note that when the force is 
calculated from the energy, both position dependences
in $G_{ii}({\bf r}, {\bf r}' = {\bf r}; \omega )$ get differentiated.
Close to a planar surface, 
both give the same contribution, since in the Weyl plane-wave
expansion (see \ref{a:Weyl}), the reflected part of the Green
tensor is proportional to ${\rm e}^{{\rm i}\gamma(z+z')}$
where $\gamma$ is the normal component of the wave vector.
In the planar geometry we consider here, the force is oriented 
perpendicular to the interface, of course.

We finally note that it is \emph{a priori} not clear that the light
force derives from a potential because we are dealing with a
dissipative system. 
At finite temperature and in equilibrium, the force
is related to the energy change when the particle is displaced. 
This creates electromagnetic fields that do work on the polarization 
fluctuations, which is partly dissipated in the system. The 
corresponding potential energy is given by the free energy of
the system. More details can be found in~\cite{Barton97} and
references therein.
Here, at zero temperature and at lowest order 
in the polarizability,
we do find a potential since ${\bf F}$ can be written as a gradient.
This need not be true at higher orders, however, because the combined 
state of atom and field develops correlations 
(the density matrix contains a term $\exp( {\bf d} \cdot 
{\bf E}( {\bf r} ) / T )$ involving the interaction Hamiltonian).

\end{small}

\subsection{Fluctuation--dissipation theorem}

We outline here a proof of the fluctuation--dissipation theorem for
the electromagnetic field close to some arbitary macroscopic
object. The only requirement is that the object interacts in a
linear way with the field and that thermal equilibrium prevails.
This text is adapted from the review paper ``Electromagnetic field 
fluctuations on the nanometer scale'' by C.\ Henkel,
submitted for publication in 
the {\it Handbook of Theoretical and Computational Nanotechnology} 
(2004).

\smallskip

\begin{small}

The fluctuations of the electromagnetic fields can be characterized by
statistical electrodynamics, which results from the application of
equilibrium thermodynamics and quantum theory to the macroscopic
Maxwell equations~\cite{Rytov3}.  We introduce in this Section the
basic definitions for the field fluctuation spectra and review how they
can be calculated.  A key result in this context is the fluctuation
dissipation theorem derived by Callen and Welton~\cite{Callen51} whose
proof is sketched here.  We conclude with some remarks on how to handle
non-equilibrium situations.

\subsubsection{Basic definition}


We shall assume that at thermodynamic equilibrium at temperature $T$,
the field and the solid medium can be 
described by a Gibbs ensemble: each state with energy $E$ 
is weighted with the Boltzmann factor $\exp( - E / k_{B} T ) \equiv
{\rm e}^{- \beta E }$. In the quantized version of the theory, these 
states are stationary states: they are eigenstates with energy $E$ of 
the corresponding Hamilton operator $\hat{H}$. We shall work in the 
Heisenberg picture where the field variables (called observables) 
evolve in time, while the state of the system is fixed. The Gibbs 
ensemble is then characterized by the density operator
\begin{equation}
    \hat \rho_{\rm eq} = \frac{ \exp( - \hat{H} / k_{B} T ) }{ 
    {\rm Tr} \,
    \exp( - \hat{H} / k_{B} T ) }
    .
    \label{eq:canonical-density-operator}
\end{equation}
This is an operator on the Hilbert space of the system that can be 
represented by a (infinite-dimensional) matrix, the density matrix.  
${\rm Tr} \, \exp( - \hat{H} / k_{B} T ) $
is the trace of the operator-valued Boltzmann factor, 
it is also called the partition function~\cite{Landau5}.

With respect to this equilibrium ensemble, we can define average 
values for the observables of interest. This
average combines the quantum expectation value in a given stationary state 
with the corresponding statistical ensemble weights. The average 
electric field, for example is given by
\begin{equation}
    \langle {\bf E}( {\bf x}, t ) \rangle
    \equiv 
  {\rm Tr}\left[ {\bf E}( {\bf x}, t ) \hat \rho_{\rm eq} \right]
=    {\rm Tr}\left[ \hat \rho_{\rm eq} {\bf E}( {\bf x}, t ) \right]
,
\label{eq:ave-el-field}
\end{equation}
where we have made use of the cyclic permutation under the trace.
The time-dependence of the field is generated by the Hamilton 
operator, so that we have
\begin{equation}
\langle {\bf E}( {\bf x}, t ) \rangle
= {\rm Tr}\left[ 
\exp( {\rm i}\hat{H} t / \hbar )
{\bf E}( {\bf x} ) 
\exp( - {\rm i}\hat{H} t / \hbar )
\hat \rho_{\rm eq} \right]
= {\rm Tr}\left[ 
{\bf E}( {\bf x} ) \hat \rho_{\rm eq} 
\right]
,
\label{eq:ave-el-field-2}
\end{equation}
where ${\bf E}( {\bf x} )$ is the electric field operator at time 
zero and we have used that the Gibbs density 
operator~(\ref{eq:canonical-density-operator})
is invariant 
under time evolution. The average can now be computed in the 
Schr\"{o}dinger picture and is found to vanish at equilibrium. In the classical 
theory, this is because the phase of the field is uniformly 
distributed. In the quantum theory, 
the stationary states for
each mode of the field (labelled by $\kappa$)
are eigenstates of the photon number operator $a^\dag_{\kappa} 
a_{\kappa}^{\phantom\dag}$. The field observable is a linear
combination of the annihilation and creation operators 
$a^\dag_{\kappa}$ and $a_{\kappa}$ that lower or raise the photon number:
their quantum expectation values thus vanish in a photon number 
eigenstate. More details can be found in Section~\ref{s:bb-spectrum}
and in the textbooks by Loudon~\cite{Loudon} and Mandel and 
Wolf~\cite{MandelWolf}.

The relevant information about the field fluctuations is thus encoded 
in the correlation function
\begin{equation}
    \langle {\bf E}( {\bf x}, t ) {\bf E}( {\bf x}', t' )
    \rangle \equiv
    {\rm Tr}\left[  
    {\bf E}( {\bf x}, t ) {\bf E}( {\bf x}', t' ) \hat\rho_{\rm eq}
    \right]
    =
    \langle {\bf E}( {\bf x}, 0 ) {\bf E}( {\bf x}', t'-t )
    \rangle
.
\label{eq:def-corr-function}
\end{equation}
In the second step, we have shifted the time arguments using the fact
that the time evolution commutes with the equilibrium density 
operator~(\ref{eq:canonical-density-operator}).
As expected from stationarity, 
this correlation function only depends on the time difference 
$\tau = t' - t$. In the limit $|\tau| \to \infty$, one expects the 
fields ${\bf E}( {\bf x}, t )$ and ${\bf E}( {\bf x}', t' )$ 
to decorrelate and the correlation function~(\ref{eq:def-corr-function})
to vanish. The time scale on which this happens gives the correlation 
or coherence time of the field. 

The spectrum of the field fluctuations can be defined by the 
Fourier expansion of the correlation function%
\footnote{%
Boldface vectors placed near each other mean a tensor product, often 
written $\otimes$. Tensors are written in calligraphic letters like 
${\cal E}$ or ${\cal H}$.}
\begin{equation}
    \langle {\bf E}( {\bf x}, t ) {\bf E}( {\bf x}', t' )
    \rangle =
    \int\limits_{-\infty}^{+\infty}\!\frac{ {\rm d}\omega }{ 2\pi }
    {\rm e}^{ - {\rm i} \omega (t' - t ) }
    \tens{E}( {\bf x}, {\bf x}'; \omega )
    .
    \label{eq:def-fluctuation-spectrum}
\end{equation}
This relation is also known as the Wiener-Khintchine 
theorem~\cite{MandelWolf}. 
The spectrum is actually a tensor, the cross-spectral 
density tensor. 
For ${\bf x} = {\bf x}'$, its components 
specify the local polarization state of the field. As a function
of ${\bf x} - {\bf x}'$, it characterizes the field's spatial 
coherence, i.e., the contrast of interference fringes in a 
double-slit experiment with slits placed by ${\bf x}$ and ${\bf x}'$,
see, e.g., the textbooks 
by Mandel and Wolf~\cite{MandelWolf} and
by Goodman~\cite{Goodman}.
As a function of the frequency $\omega$, the spectrum 
specifies the strength of the field fluctuations. This can 
be seen from
the equivalent relation for the Fourier transforms of the field
\begin{equation}
    \langle {\bf E}( {\bf x}, -\omega ) {\bf E}( {\bf x}', \omega' )
    \rangle =
2\pi\delta( \omega - \omega' )
\,\tens{E}( {\bf x}, {\bf x}'; \omega' )
=
\langle [ {\bf E}( {\bf x}, \omega ) ]^\dag 
{\bf E}( {\bf x}', \omega' ) \rangle
.
\label{eq:freq-correlations}
\end{equation}
In the last step, we have made use of the reality 
of the electric field that leads to identity
\(
    [{\bf E}( {\bf x}, \omega ) ]^\dag =
    {\bf E}( {\bf x}, -\omega )
\)
for real $\omega$. This way of writing also 
shows that the diagonal elements of $\tens{E}( {\bf x}, {\bf x}; 
\omega )$ are positive. More generally, 
$0 \le {\bf u}^* \cdot \tens{E}( {\bf x}, {\bf x}; 
\omega ) \cdot {\bf u}$ for any complex vector ${\bf u}$.
We note that the Fourier transforms of the fields strictly speaking 
do not exist as ordinary functions. The Fourier calculus nevertheless 
applies symbolically for the corresponding operator-valued 
distributions.

The fluctuation spectrum of the electric field plays a key role for 
spontaneous and stimulated decay on the electric dipole transitions 
of an atom or molecule. The corresponding spectrum for the magnetic 
field characterizes the perturbation the field exerts on an atomic
magnetic moment or spin. This is discussed in detail in 
Section~\ref{s:atom-chips}.

\subsubsection{Blackbody fluctuations}
\label{s:bb-spectrum}

Let us illustrate the correlation function introduced above with the 
example of the blackbody radiation field in free space. 
The electric field operator 
can in that case be expanded in plane wave modes~\cite{Loudon,MandelWolf}
\begin{equation}
    {\bf E}( {\bf x}, t ) =
    \sum\limits_{ \kappa }
    \sqrt{ \frac{ \hbar \omega( k ) }{ 2\varepsilon_{0} V} }
    \left[ a_{\kappa}( t ) \beps 
    \exp( {\rm i} {\bf k} \cdot {\bf x} )
    + \textsc{h.c.}
    \right]
    ,
    \label{eq:el-mode-expansion}
\end{equation}
where the mode label $\kappa = ({\bf k},\,\beps)$ combines the
wave vector ${\bf k}$ and the polarization vector 
$\beps\perp {\bf k}$,
and $V$ is the quantization volume. 
We assume periodic boundary 
conditions so that the allowed wave vectors are discrete.
The mode frequency is given by $\omega( k ) = c k = 
c \sqrt{ {\bf k}^2 }$.
`\textsc{h.c.}' denotes the hermitean conjugate operator so that the 
electric field is globally hermitean.
In the absence of any sources, 
the annihilation and creation operators evolve according to
\begin{equation}
    a_{\kappa}( t ) = a_{ \kappa } 
    \exp[ - {\rm i} \omega( k )t  ]
    , \qquad
    a_{ \kappa }^\dag( t ) = a_{ \kappa }^\dag 
    \exp[ {\rm i} \omega( k )t  ]
.
\label{eq:a-of-time}
\end{equation}
The Schr\"{o}dinger operators
$a_{ \kappa }$ and $a_{ \kappa }^\dag$
satisfy the bosonic commutation relations
\begin{equation}
    \left[ 
    a_{ \kappa }^{\phantom\dag}, \,a_{ \kappa' }^\dag
    \right]
    \equiv
    a_{ \kappa }^{\phantom\dag} \,a_{ \kappa' }^\dag
-
a_{ \kappa' }^\dag\,a_{ \kappa }^{\phantom\dag} 
=
\delta_{\kappa, \, \kappa'} \equiv
\delta_{{\bf k},\,{\bf k}'} 
\delta_{\beps,\,\beps'}
.
    \label{eq:a-adag-commutator}
\end{equation}
The equilibrium expectation value of products of the 
mode operators decorrelates for different modes 
because the
density operator factorizes into a product over all modes.
Therefore, the expectation value $\langle 
a_{ \kappa' }^\dag\,a_{ \kappa }^{\phantom\dag} \rangle$
vanishes for $\kappa \ne \kappa'$.

For a given mode $\kappa$, one gets from the
Gibbs ensemble the Bose-Einstein occupation number
(with $\beta \equiv 1/(k_{\rm B} T)$)
\begin{eqnarray}
    \langle a_{\kappa}^{\dag} a_{\kappa}^{\phantom\dag} \rangle
    &=& \bar{n}[ \omega( k ) ] \equiv
    \frac{ 1 }{ {\rm e}^{ \beta \hbar \omega( k ) } - 1 }
    \label{eq:Bose-Einstein-occupation-1}
    \\
    \langle a_{\kappa}^{\phantom\dag} a_{\kappa}^{\dag} \rangle
    &=& 1 + \bar{n}[ \omega( k ) ] =
    \frac{ 1 }{ 1 - {\rm e}^{ - \beta \hbar \omega( k ) } }
,
\label{eq:Bose-Einstein-occupation}
\end{eqnarray}
while the products $a_{\kappa} a_{\kappa}$ and 
$a_{\kappa}^{\dag} a_{\kappa}^{\dag}$ have zero average.
To prove Eq.~(\ref{eq:Bose-Einstein-occupation-1})
, we evaluate
the trace in the eigenbasis $| n_{\kappa} \rangle$
of the photon number operator
$a^\dag_{\kappa} a_{\kappa}^{\phantom\dag}$
for the given mode. The energy eigenvalue of 
$| n_{\kappa} \rangle$ is $\hbar \omega(k) (n_{\kappa} + \frac12)$,
and the summation over the Boltzmann weights gives
\begin{eqnarray}
    \langle a_{\kappa}^{\dag} a_{\kappa}^{\phantom\dag} \rangle
    & = & 
    \frac{ \sum\limits_{n_{\kappa} = 0}^{\infty} n_{\kappa} 
    {\rm e}^{ - \beta \hbar \omega(k) (n_{\kappa} + \frac12) } }
    { \sum\limits_{n_{\kappa} = 0}^{\infty} 
	{\rm e}^{ - \beta \hbar \omega(k) (n_{\kappa} + \frac12) } }
    \nonumber\\
    & = & -\frac{ \partial }{ \partial \xi }
    \left.
    \log\left( \sum\limits_{n = 0}^{\infty} 
    {\rm e}^{ - \xi n } \right)
    \right|_{\xi = \beta\hbar\omega(k)}
    =
    \frac{ \partial }{ \partial \xi }
	\left.
	\log\left( 1 - {\rm e}^{ - \xi } \right)
	\right|_{\xi = \beta\hbar\omega(k)}
    \nonumber\\
    & = & 
    \frac{ 1 }{ {\rm e}^{ \beta \hbar \omega( k ) } - 1 }
    .
    \label{eq:proof-Bose-Einstein-occupation}
\end{eqnarray}

Using these informations, a straightforward calculation in the 
continuum limit,
$\sum_{{\bf k}} \mapsto V \int{\rm d}^3k / (2\pi)^3$,
leads to
\begin{eqnarray}
    \tens{E}( {\bf x}, {\bf x} + {\bf r}; \omega )
    & = &
    \frac{ \hbar \omega^3 \bar n( \omega ) 
    }{ 2\pi\varepsilon_{0} c^3 } 
     \left\{
     (\mathbbm{ 1 } - \hat{\bf r} \hat{\bf r}) \frac{ \sin (\omega r / 
     c) }{ \omega r / c }
     \right.
     \label{eq:blackbody-e-correlation}
     \\
     && \qquad 
     \left. {} +
     (\mathbbm{ 1 } - 3 \hat{\bf r} \hat{\bf r}) 
     \left( \frac{ \cos(\omega r / c) }{ (\omega r / c)^2 }
     -
     \frac{ \sin(\omega r / c) }{ (\omega r / c)^3 }
     \right) \right\}
     ,
     \nonumber
\end{eqnarray}
which only depends on the difference vector ${\bf r} = {\bf x}' - {\bf 
x}$, as expected. This expression is very similar to the Green 
tensor in free space
which is not a coincidence, but
a special case of the fluctuation-dissipation theorem introduced
in Section~\ref{s:FD-theorem}.

In the limit ${\bf r} \to {\bf 0}$, one can check that $\tens{E}( {\bf
x}, {\bf x}; \omega )$ is proportional to the unit tensor and positive.
Taking the trace, one gets, up to a factor $\varepsilon_{0}/2$, the
spectrum of the electric energy density $u_{\rm e}( \omega )$.  Summing
the contributions of positive and negative frequencies, the electric
energy density is given by the Planck formula
\begin{eqnarray}
U_{\rm bb,\,e} &=& \int\limits_0^{\infty}\!\frac{ {\rm d}\omega }{ 2\pi } \,
u_{\rm bb,\,e}( \omega ) 
,
\label{eq:integral-el-energy-density}
\\
u_{\rm bb,\,e}( \omega )  &=&
\frac{ \hbar \omega^3 ( \bar n( \omega ) + \frac12 ) }
{ \pi c^3 }
=
2\pi \textsc{dos}( \omega ) \frac{ \hbar\omega }{ 2 }
\left(  \bar n( \omega ) + \frac12   \right)
.
\end{eqnarray}
In the last expression, we have made use of the free space 
\textsc{dos}$(\omega) = \omega^2 / \pi^2 c^3$. This result has an intuitive
explanation: the electric energy density is 
the density of modes per unit volume,
$\textsc{dos}( \omega ) {\rm d}\omega$,
multiplied by one 
half of the average equilibrium energy $\hbar\omega
\big(  \bar n( \omega ) + \frac12   \big)$ 
of a harmonic oscillator (the other half contributes to the
magnetic energy). 

At zero temperature, the electric plus
magnetic energy per mode is given by the ground state 
oscillator energy $\hbar\omega/2$, and this leads to a divergent 
integral in the \textsc{uv}. 
From this divergent zero-point energy, one can extract 
a finite, measurable energy difference,
called the Casimir energy,
when the mode functions are changed by the boundary conditions
imposed by material structures. See 
the textbook by Mostepanenko and Trunov~\cite{Mostepanenko}
for more details. The energy measured by a photodetector does not
diverge because it is proportional to average of the 
photon number operator 
$\langle a_\kappa^\dag a_\kappa^{\phantom\dag}\rangle = \bar n( 
\omega(k))$~\cite{Loudon,MandelWolf}.
If the creation and annihilation operators are ordered in this way, 
the divergent contribution of the zero-point energy disappears.
The blackbody spectrum then shows an exponential
decrease beyond the thermal wavelength 
$\lambda_{\rm th} = 2\pi\hbar c / (k_{\rm B} T)$ 
(the Wien displacement law), and the 
frequency integral~(\ref{eq:integral-el-energy-density}) 
becomes convergent.

\subsubsection{Fluctuation-dissipation theorem}
\label{s:FD-theorem}

The previous example suggests that there is a relation between the
field fluctuation spectrum and the Green function. With the previously 
introduced notation,
we have
\begin{equation}
\tens{E}( {\bf x}, {\bf x}'; \omega )
= \frac{ 2\hbar }{ {\rm e}^{ \beta \hbar \omega } - 1 }
{\rm Im}\,\tens{G}( {\bf x}, {\bf x}'; \omega )
,
\label{eq:FD-theorem}
\end{equation}
where the Green tensor is defined as the electric field radiated by a 
monochromatic point dipole,
\begin{equation}
    {E}_{i}^{({\rm dip})}( {\bf x}; \omega ) = 
    {\cal G}_{ij}( {\bf x}, {\bf r}; \omega ) d_{j}( \omega )
    ,
\label{eq:def-Green}
\end{equation}
and the fluctuation spectrum by Eq.~(\ref{eq:def-fluctuation-spectrum}).
Eq.~(\ref{eq:FD-theorem}) is actually true under more general conditions
and is known as a fluctuation-dissipation (FD) 
theorem~\cite{Callen51}. It holds for linear
systems and their fluctuations around the thermal equilibrium state.
The dissipation is encoded in the imaginary part of the response 
function that characterizes the linear response of the system to an external
perturbation. The FD theorem is of the 
form given here provided the Green tensor satisfies the symmetry 
condition specified in Eq.~(\ref{eq:symmetry-Green}).

The FD theorem will be our basic tool to compute field fluctuation
spectra in the near field of nanostructures. We give in this Section
an introduction for linear systems and summarize a general
proof in the context of statistical electrodynamics. A generalization
to non-equilibrium situations is discussed in 
Section~\ref{s:non-equilibrium}.

\paragraph{Johnson-Nyquist noise in metals.}
%
Consider a resistance at temperature $T$. One observes
a thermal fluctuation of the current through the resistance, 
called Johnson noise, whose
variance in a given bandwidth $\Delta f = \Delta \omega / (2\pi)$
is given by the Nyquist formula
\begin{equation}
\langle \delta I^2 \rangle_{\Delta\omega} =
\frac{ \Delta \omega }{ 2 \pi }
\frac{ 4 k_{\rm B} T }{ R( \omega ) }
.
\label{eq:Nyquist-1}
\end{equation}
We now show that this formula can be related to a fluctuation
dissipation theorem for the current density ${\bf j}( {\bf x} )$
of the resistance. Consider first the
$z$-component of the current density and a 
small volume element $\Delta V = \Delta z\,\Delta A$. From the
current noise along the $z$-direction, one then has
\begin{equation}
\langle \delta j_z( {\bf x} )^2 \rangle_{\Delta\omega} =
\frac{ \Delta \omega }{ 2 \pi }
\frac{ 4 k_{\rm B} T }{ R( \omega ) \Delta A^2 }
=
\frac{ \Delta \omega }{ 2 \pi }
\frac{ 4 k_{\rm B} T \sigma( \omega ) }{ \Delta V }
,
\label{eq:Nyquist-2}
\end{equation}
where $\sigma( \omega ) = \Delta z / ( R( \omega) \Delta A)$ is the 
conductivity (the inverse of the specific resistance). 
In the low frequency range where the Nyquist formula is
valid, the conductivity can be expressed via the dielectric function
of the resistance, $\varepsilon_0 \varepsilon( \omega ) = \varepsilon_0 
+ {\rm i}
\sigma / \omega$. 
The result~(\ref{eq:Nyquist-2}) can thus be obtained by averaging 
the following relation over the volume element
\begin{equation}
\langle \delta {\bf j}( {\bf x} ) \delta {\bf j}( {\bf x}' ) 
\rangle_{\Delta\omega} =
\frac{ \Delta \omega }{ 2 \pi }
4 k_{\rm B} T \omega\,
\mathbbm{1} {\rm Im}\,\varepsilon_0 \varepsilon( {\bf x}; \omega )
\delta( {\bf x} - {\bf x}' )
.
\label{eq:Nyquist-3}
\end{equation}
We have assumed that neighboring volume elements have uncorrelated 
current noise, hence the spatial delta function.
We use here the convention that
the current noise~(\ref{eq:Nyquist-3}) is given by the integral 
of the noise spectrum
$\tens{J}( {\bf x}, {\bf x}'; \omega )$
over intervals $\Delta\omega / (2\pi)$ centered at 
positive and negative frequencies
$\pm\omega$. Since $\omega\,{\rm Im}\,\varepsilon( {\bf x}; \omega )$ is
an even function of $\omega$, the noise spectrum is given by
\begin{equation}
\tens{J}( {\bf x}, {\bf x}'; \omega )
=
2 k_{\rm B} T \omega
\,\mathbbm{1} \,
{\rm Im}\,\varepsilon_0 \varepsilon( {\bf x}; \omega )
\delta( {\bf x} - {\bf x}' )
.
\label{eq:Nyquist-3b}
\end{equation}
This result has already the structure of the FD theorem~(\ref{eq:FD-theorem}).
Since the dielectric function gives the polarization induced
by an electric field, a more natural formulation is in terms of
the polarization noise spectrum (writing ${\bf j}( \omega ) 
= - {\rm i} \omega {\bf P}( \omega )$)
\begin{equation}
\tens{P}( {\bf x}, {\bf x}'; \omega )
=
\frac{ 2 k_{\rm B} T }{ \omega } 
\,\mathbbm{1} \,
{\rm Im}\,\varepsilon_0 \varepsilon( {\bf x}; \omega )
\delta( {\bf x} - {\bf x}' )
.
\label{eq:Nyquist-4}
\end{equation}
The temperature-dependent prefactor is the low-frequency limit of 
$2\hbar / ({\rm e}^{ \hbar\omega/k_{\rm B} T } - 1 )$ occurring in
Eq.~(\ref{eq:FD-theorem}). The spatial $\delta$-function applies to
a local dielectric response. The generalization
to a nonlocal medium is immediate: the fluctuations are then correlated 
on some characteristic scale, typically the mean free path. 

We thus find that the strength of the thermal Johnson noise at a given
frequency in an absorbing material is related to the amount of
dissipation, as encoded in the imaginary part ${\rm Im}\,\varepsilon(
\omega )$.  This permits to characterize the thermal polarization and 
magnetization fluctuations that appear in the macroscopic
Maxwell equations.
The noise spectrum of the polarization
noise ${\bf P}_{\rm fl}( {\bf x}; \omega )$ is given by the
Johnson-Nyquist formula~(\ref{eq:Nyquist-4}), with the factor $k_{\rm
B} T$ replaced by $\hbar \omega / ( {\rm e}^{\beta \hbar \omega } - 1
)$ to be valid at all frequencies.  If the material is magnetic with a
(local) permeability $\mu( {\bf x}; \omega )$, it contains
magnetization fluctuations ${\bf M}_{\rm fl}( {\bf x}; \omega )$ with a
spectrum
\begin{equation}
    \tens{M}( {\bf x}, {\bf x}; \omega ) = -
    \frac{ 2 \hbar \,\mathbbm{1} }{ {\rm e}^{\beta \hbar \omega } - 1 }
    {\rm Im} \frac{1 }{ \mu_{0} \mu( {\bf x}; \omega ) }
    \delta( {\bf x} - {\bf x}' )
    .
    \label{eq:FD-magnetization}
\end{equation}
The fluctuating material polarization radiates an electromagnetic field
that, in thermodynamic equilibrium, compensates for the loss of
electromagnetic energy inside the material. Only in this way
is it possible to enforce the equipartition law of thermal equilibrium,
every degree of freedom (here the polarization field) carrying an
energy $k_B T / 2$. Consider the balance of electromagnetic energy 
for a system without external 
polarization sources. On the one hand, the mechanical work performed 
per unit time can be written as
\begin{eqnarray}
&&
\omega \,{\rm Im}\langle  {\bf P}_{\rm fl}^\dag( {\bf x}; \omega )
    \cdot {\bf E}( {\bf x}; \omega' ) \rangle
    \nonumber\\
    &&= 
2\pi \omega \delta( \omega - \omega' )
\int\!{\rm d}^3 x' \,
\tens{P}_{ij}( {\bf x}, {\bf x}'; \omega )
\,{\rm Im}\,\tens{G}_{ij}( {\bf x}, {\bf x}'; \omega ) 
\nonumber\\
&&=
2\pi \delta( \omega - \omega' )
\frac{ 2 \hbar \omega }{ {\rm e}^{\beta \hbar \omega } - 1 }
{\rm Im}\,\varepsilon_{0} \varepsilon( {\bf x}; \omega ) 
\,{\rm Im}\,{\rm Tr}\,\tens{G}( {\bf x}, {\bf x}; \omega ) 
.
    \label{eq:ave-PdotE}
\end{eqnarray}
We have taken into account that only the part of the field 
radiated by the polarization fluctuation is correlated with 
the latter and expressed that field in terms of the Green 
tensor~(\ref{eq:def-Green}). A similar result holds for the
magnetic contribution. On the other hand, the average over the 
electric losses in the medium leads to the same expression
\begin{eqnarray}
&&
\omega
\,{\rm Im}[\varepsilon_{0} \varepsilon( {\bf x}; \omega ) ]
\langle  {\bf E}^\dag( {\bf x}; \omega )
    \cdot {\bf E}( {\bf x}; \omega' ) \rangle
    \nonumber\\
    &&= 
2\pi \delta( \omega - \omega' )
\,{\rm Im}[\varepsilon_{0} \varepsilon( {\bf x}; \omega ) ]
\frac{ 2 \hbar \omega }{ {\rm e}^{\beta \hbar \omega } - 1 }
\,{\rm Im}\,{\rm Tr}\,\tens{G}( {\bf x}, {\bf x}; \omega ) 
,
    \label{eq:ave-EdotE}
\end{eqnarray}
using the fluctuation-dissipation theorem~(\ref{eq:FD-theorem}).  As a
result, the energy the polarization emits into the field
[Eq.(\ref{eq:ave-PdotE})] is exactly compensated for by the field
energy lost by absorption, Eq.(\ref{eq:ave-EdotE}). 
This also implies that
the average Poynting vector ${\rm Re}\langle {\bf E}^\dag \times {\bf H}
\rangle$ vanishes, since there is no net energy transfer between medium
and field.

We shall see that in the quantized theory, the
polarization fluctuations of the material also contribute to the
quantum fluctuations of the field.  Otherwise, the dissipation present
in the macroscopic Maxwell equations would force the field operators to
decay to zero, including their commutators.  These are preserved due to
the quantum fluctuations of the material polarization.  This picture
suggests as well the existence of an FD theorem: the material loss that
forces the fields to decay must be balanced by the fluctuations inside
the material.

\paragraph{Properties of quantum field fluctuations.}
%
Before giving a general proof of the FD theorem~(\ref{eq:FD-theorem}),
let us summarize some of the properties it implies for the equilibrium
fluctuations of quantized fields.

The FD theorem allows to compute the quantum and
thermal fluctuations of the electromagnetic field once the Green 
tensor is known. This quantity can be computed by solving the 
macroscopic Maxwell equations with point-like dipole sources.
As long as the medium responds linearly to the field,
the classical version of the theory is sufficient, quantum and
thermal fluctuations are handled self-consistently using the theorem.

The noise spectrum of a quantized field is not symmetric. It is 
proportional to the Bose-Einstein occupation number $\bar n( \omega )$
for positive frequencies and decays to zero for 
$\hbar\omega \gg k_{\rm B}T$.
At negative frequencies, one finds,
given that ${\rm Im}\,\tens{G}(\omega)$ is
an odd function,
that the spectrum is proportional 
to $1 + \bar n( |\omega| )$.
The zero-point fluctuations appear here. The asymmetric frequency 
spectrum of zero-point or vacuum fluctuations
can be understood qualitatively by noting that in the
ground state, a system can only fluctuate via a virtual transition
towards a state with higher energy. The corresponding Bohr 
frequencies are all positive. (That this leads to a spectral weight
at negative $\omega$ is related to our -- conventional -- choice
of the exponential factor in the noise 
spectrum~(\ref{eq:def-fluctuation-spectrum}).)
In the high-temperature limit or, 
equivalently, for classical systems, upward and downward transitions 
occur with equal probability, and the fluctuation spectrum is 
symmetric: $\bar n( \omega ) \approx \bar n( |\omega| ) + 1
\approx k_{\rm B} T / \hbar\omega \gg 1$. For intermediate 
temperatures, we show below that the principle of detailed balance is 
satisfied, up- and downward transition rates differing by a 
factor ${\rm e}^{\beta\hbar\omega}$.

At positive frequencies, Eq.~(\ref{eq:def-fluctuation-spectrum}) 
shows
that the spectrum $\tens{E}({\bf x}, {\bf x}'; \omega)$
picks out that part ${\bf E}^{(+)}({\bf x}', t')$ of the 
electric field operator that evolves 
like ${\rm e}^{- {\rm i} \omega t'}$. 
By analogy to time-dependent wave functions in quantum mechanics,
this part is called the positive frequency part of the field.
In the mode 
expansion~(\ref{eq:el-mode-expansion}), it corresponds to the sum
over the annihilation operators $a_{\kappa}$ [see also 
Eq.~(\ref{eq:a-of-time})]. 
Similarly, only the negative frequency part
${\bf E}^{(-)}({\bf x}, t) = [{\bf E}^{(+)}({\bf x}, t)]^\dag$
of the field operator contributes in the first factor.
It follows that in the vacuum state, 
the expectation value
\begin{equation}
    {\rm Tr}\, \left[
    {\bf E}^{(-)}({\bf x}, t) {\bf E}^{(+)}({\bf x}', t')
    \hat\rho_{\rm vac} \right]
    = 0
    \label{eq:normal-order}
\end{equation}
vanishes since the annihilation operators, by definition,
give zero when acting on the vacuum state. 
This operator order (annihilation operators acting first)
is usually called `normal' order.
A typical example is the intensity measured by a 
photodetector~\cite{Loudon,MandelWolf}. 
A nonzero vacuum expectation value occurs with the reverse operator order
(creation operators acting first). This anti-normal order is 
picked out for negative frequencies in the fluctuation 
spectrum $\tens{E}({\bf x}, {\bf x}'; \omega)$. It gives 
nonzero results even in the vacuum state (at zero temperature),
as we have seen in Eq.~(\ref{eq:Bose-Einstein-occupation}) and
in the FD theorem~(\ref{eq:FD-theorem}).

\paragraph{Proof of the FD theorem with linear response theory.}
%
%
As mentioned previously, the macroscopic Maxwell
equations in an absorbing medium have to be supplemented by  
material fluctuations in order to be consistent with thermodynamics 
and quantum theory. We thus 
split the polarization and magnetization operators into
\begin{eqnarray}
{\bf P}( {\bf x}, t ) &\mapsto&
{\bf P}_{\rm fl}( {\bf x}, t ) + {\bf P}_{\rm ext}( {\bf x}, t )
\nonumber\\
{\bf M}( {\bf x}, t ) &\mapsto&
{\bf M}_{\rm fl}( {\bf x}, t ) + {\bf M}_{\rm ext}( {\bf x}, t )
\label{eq:add-q-noise}
\end{eqnarray}
where the terms with the subscript `fl' describe the fluctuations in the
material and the `ext' all other sources like the dipole moments of
atoms or molecules. 
In thermal equilibrium, the fluctuations average
to zero, as we found after Eq.~(\ref{eq:ave-el-field-2}). 
In the following, more explicit information about the polarization noise
is not needed. We shall assume that a Hamilton operator $\hat H$ 
exists that
generates the macroscopic Maxwell equations as the Heisenberg equations
of motion for the electric and magnetic field operators. 
(We are actually
adopting a quantum Langevin picture, see Mandel and Wolf~\cite{MandelWolf}
and Gardiner~\cite{Gardiner}.)
A similar demonstration has been given by Wylie and Sipe~\cite{Sipe84}.


The field fluctuation spectrum, from 
Eq.~(\ref{eq:def-fluctuation-spectrum}), is given by the expectation 
value
\begin{equation}
    \tens{E}( {\bf x}, {\bf x}'; \omega )
    = 
    \int\limits_{-\infty}^{+\infty}\! {\rm d}\tau\,
    {\rm e}^{ {\rm i} \omega \tau }
    \langle {\bf E}( {\bf x}, 0 ) {\bf E}( {\bf x}', \tau )
    \rangle
.
\label{eq:FD-proof-spectrum}
\end{equation}
We assume thermal equilibrium without external sources and
have used the stationarity of the correlation function. We 
now connect this spectrum to the linear response of the field to an 
external dipole oscillator, following Callen and Welton~\cite{Callen51}.

The solution for the electric field operator in the presence of a 
polarization source can be represented
in terms of the Green tensor in the quantum theory as well because the 
substitution~(\ref{eq:add-q-noise}) preserves the linearity of
the macroscopic Maxwell equations. 
We thus get the field due to an
operator-valued source, plus a term describing the
free evolution of the field
\begin{eqnarray}
&&     {\bf E}( {\bf x}, t ) =
    {\bf E}_{\rm free}( {\bf x}, t )
\nonumber\\
&& {}
+ \int\!\frac{ {\rm d}\omega }{ 2 \pi }
{\rm e}^{ - {\rm i} \omega t } 
    \int\limits_{V}\!{\rm d}^3x' \,
    \tens{G}( {\bf x}, {\bf x}'; \omega ) 
    \cdot \left[
{\bf P}_{\rm fl}( {\bf x}'; \omega ) + {\bf P}_{\rm ext}( {\bf x}'; \omega )
\right]
    \label{eq:q-field-t}
.
\end{eqnarray}
There is a similar contribution from the magnetization that we do not 
need for the present discussion. 
In equilibrium, the free field operator has zero average, and 
we get the expectation value 
\begin{equation}
    \langle {\bf E}( {\bf x}, t ) \rangle =
\int\!\frac{ {\rm d}\omega }{ 2 \pi }
{\rm e}^{ - {\rm i} \omega t } 
    \int\limits_{V}\!{\rm d}^3x' \,
    \tens{G}( {\bf x}, {\bf x}'; \omega ) 
    \cdot 
\langle
{\bf P}_{\rm ext}( {\bf x}'; \omega )
\rangle
    \label{eq:ave-q-field-t-2}
.
\end{equation}
The Green tensor can thus be identified with the linear response of the
average field to a classical external polarization source (where
${\bf P}_{\rm ext}( {\bf x}'; \omega )$ is c-number valued).


The linear response of the field can also be calculated directly
from the Heisenberg equations. This provides us with an alternative
expression for the Green tensor where equilibrium
correlations will become apparent.
For simplicity, we focus in the following on the response to an electric
point  dipole at the position ${\bf x}'$.  
The coupling of the field to the dipole
is described by adding to the Hamiltonian the term
\begin{equation}
{H}_{\rm int}( t ) =
- {\bf d}( t ) \cdot
{\bf E}( {\bf x}', t )
,
\end{equation}
and the Heisenberg equation reads
\begin{eqnarray}
\frac{ {\rm d} }{ {\rm d}t }
{\bf E}( {\bf x}, t ) &=&
- \frac{ {\rm i} }{ \hbar }
\left[ {\bf E}( {\bf x}, t ), \, \hat{H} \right]
+
\frac{ {\rm i} }{ \hbar }
\left[ 
{\bf E}( {\bf x}, t ) 
,\, 
{E}_j( {\bf x}', t ) 
\right]
{d}_j( t ),
\end{eqnarray}
where summation over $j$ is understood in the last term.
The first term generates the free evolution of the field.
Solving to first order in ${\bf d}$ and taking the average,
we identify the field response function in the time domain as
\begin{eqnarray}
\langle
{E}_{i}( {\bf x}, t ) \rangle 
&=&
\int\limits_{-\infty}^{+\infty}
\!{\rm d}\tau \, 
\chi_{ij}( {\bf x}, {\bf x}', \tau )
{d}_{j}( t - \tau )
\\
\chi_{ij}( {\bf x}, {\bf x}', \tau )
&=&
\left\{
\begin{array}{ll}
{\displaystyle \frac{ {\rm i} }{ \hbar } }
\langle
\left[ 
{E}_{i}( {\bf x}, t ) 
,\, 
{E}_j( {\bf x}', t - \tau ) 
\right]\rangle
& \mbox{for } \tau \ge 0,
\\
0 & \mbox{for } \tau < 0,
\end{array}
\right.
\label{eq:def-response-function-chi}
\end{eqnarray}
where the time dependence of the field operators is that of 
the evolution under $\hat H$. 
The response function is thus itself a correlation function 
of the field.
Due to the stationarity of equilibrium, 
Eq.~(\ref{eq:def-response-function-chi})
does not depend on $t$. 
It can be checked directly that 
$\chi_{ij}( {\bf x}, {\bf x}', \tau )$ is real as in the classical 
theory. 
Taking the Fourier transform of $\chi_{ij}$,
we thus get an expression for the Green tensor in terms of
a field correlation spectrum
\begin{equation}
    \tens{G}_{ij}( {\bf x}, {\bf x}'; \omega )
    = 
    \frac{ {\rm i} }{ \hbar }
    \int\limits_{0}^{+\infty}
    \!{\rm d}\tau \, {\rm e}^{ {\rm i} \omega \tau}
    \langle
    \left[ 
    {E}_{i}( {\bf x}, \tau ) 
    ,\, 
    {E}_j( {\bf x}', 0 ) 
    \right]\rangle
    \label{eq:Green-and-commutator}
.
\end{equation}
By causality, the time integral is running over one half of the 
real axis only. We can make an integral over all $\tau$ appear, as
it occurs in the fluctuation spectrum~(\ref{eq:FD-proof-spectrum}),
by forming the combination
\begin{eqnarray}
    &&
    \frac{ 1 }{ 2 {\rm i}Ê}
    \left\{
    \tens{G}_{ji}( {\bf x}', {\bf x}; \omega )
    -
    [\tens{G}_{ij}( {\bf x}, {\bf x}'; \omega )]^{*} 
    \right\}
    = 
    \nonumber\\
    && \qquad
-   \frac{ 1 }{ 2 \hbar }
    \int\limits_{-\infty}^{+\infty}
    \!{\rm d}\tau \, 
    {\rm e}^{ {\rm i}\omega \tau}
    \langle
    \left[ 
    {E}_i( {\bf x}, 0 ) 
    ,\, 
    {E}_{j}( {\bf x}', \tau ) 
    \right]\rangle
    \label{eq:Green-and-commutator-2}
.
\end{eqnarray}
The following relation allows to permute operators occurring
in equilibrium correlation functions:
\begin{equation}
    \int\limits_{-\infty}^{+\infty}
    \!{\rm d}\tau \, 
    {\rm e}^{ {\rm i}\omega \tau}
    \langle   A( \tau )  B( 0 )\rangle
    =
    {\rm e}^{ \beta \hbar \omega} \!
    \int\limits_{-\infty}^{+\infty}
    \!{\rm d}\tau \, 
    {\rm e}^{ {\rm i}\omega \tau}
    \langle   B(0) A( \tau )  \rangle
.
\label{eq:permute-operators}
\end{equation}
In the classical theory, $\hbar = 0$ and operator ordering is 
irrelevant. 
Using this identity in the second term of 
the commutator in Eq.~(\ref{eq:Green-and-commutator-2}), we 
find the FD theorem:
\begin{eqnarray}
    \tens{E}_{ij}( {\bf x}', {\bf x}; \omega ) 
    &=&
    \frac{ 2 \hbar }{ {\rm e}^{ \beta \hbar \omega} - 1 }
    \frac{  
    \tens{G}_{ji}( {\bf x}', {\bf x}; \omega )
    -
    [\tens{G}_{ij}( {\bf x}, {\bf x}'; \omega )]^{*}
    }{ 2 {\rm i} }
.
\label{eq:FD-spectrum-wo-reciprocity}
\end{eqnarray}
The form~(\ref{eq:FD-theorem}) is recovered when the Green tensor 
satisfies the symmetry relation~(\ref{eq:symmetry-Green}). This
requires the additional assumption that permittivity and permeability
are symmetric, an assumption that we shall make in this contribution.

Eq.~(\ref{eq:permute-operators}) can be proved using
the Gibbs density operator~(\ref{eq:canonical-density-operator})
and the solution for the Heisenberg operator $A(\tau)$
[see also Eq.~(\ref{eq:ave-el-field-2})]:
\begin{equation}
    {\rm e}^{{\rm i} \omega \tau}
    \langle A( \tau )  B( 0 ) \rangle
    = {\rm e}^{{\rm i} \omega \tau}
    \frac{
    {\rm Tr}\left[
    \exp({\rm i} \hat H \tau / \hbar)
    A 
    \exp(- {\rm i} \hat H \tau / \hbar)
    B
    \exp(- \beta \hat H)
    \right]
    }{
    {\rm Tr}\,\exp(- \beta \hat H) }
    \label{eq:permute-operators-1}
\end{equation}
One shifts the integration path in the complex $\tau$-plane to the
line $-\infty - {\rm i} \hbar\beta$ $\ldots$ $+\infty - {\rm i} \hbar\beta$
and assumes that for $|\tau|\to \infty$, the correlation function
vanishes (otherwise this limiting value can be subtracted). Along the
shifted path, Eq.~(\ref{eq:permute-operators-1}) becomes
\begin{eqnarray}
    &&
    {\rm e}^{\beta \hbar \omega}
    {\rm e}^{{\rm i} \omega \tau}
    \frac{
    {\rm Tr}\left[
    \exp(\beta \hat H)
    \exp({\rm i} \hat H \tau / \hbar)
    A 
    \exp(-{\rm i} \hat H \tau / \hbar)
    \exp(- \beta \hat H)
    B
    \exp(- \beta \hat H)
    \right]
    }{
    {\rm Tr}\,\exp(- \beta \hat H) }
    \nonumber
    \\
    &&
    =
    {\rm e}^{\beta \hbar \omega}
    {\rm e}^{{\rm i} \omega \tau}
    \frac{
    {\rm Tr}\left[
    B
    \exp({\rm i} \hat H \tau / \hbar)
    A 
    \exp(-{\rm i} \hat H \tau / \hbar)
    \exp(- \beta \hat H)
    \right]
    }{
    {\rm Tr}\,\exp(- \beta \hat H) }
    \nonumber\\
    && =
    {\rm e}^{\beta \hbar \omega}
    {\rm e}^{{\rm i} \omega \tau}
    \langle B( 0 )  A( \tau ) \rangle
    ,
    \label{eq:permute-operators-2}
\end{eqnarray}
using cyclic permutation under the trace. The $\tau$-integral now 
yields the right hand side of Eq.~(\ref{eq:permute-operators}).

\subsubsection{Non-equilibrium situations}
\label{s:non-equilibrium}


A typical non-equilibrium situation that occurs in physics on the 
nanometer scale is a temperature gradient inside a nanostructure.
In thermal scanning probe microscopy, to quote another example, 
structures are held at different temperatures, being in
contact with different reservoirs. These kind of settings
can be described by a slight generalization of the present 
theory provided one assumes that each volume element of the solid 
structure is locally in thermal equilibrium at the temperature 
$T( {\bf x} )$. In this case, we can write down the fluctuation 
dissipation theorem for the thermal polarization field, by
generalizing Eq.~(\ref{eq:Nyquist-4}),
\begin{equation}
    \tens{P}( {\bf x}, {\bf x}'; \omega )
=
\frac{ 2 \hbar 
\,\mathbbm{1} \,
{\rm Im}\,\varepsilon_0 \varepsilon( {\bf x}; \omega )
}{ \exp[ \hbar\omega / k_{\rm B} T({\bf x}) ]
- 1} 
\delta( {\bf x} - {\bf x}' )
.
    \label{eq:local-FD-polarization}
\end{equation}
We have assumed a local dielectric response for simplicity.
The corresponding field fluctuation spectrum can be computed from
the field operator~(\ref{eq:q-field-t}) where the freely evolving 
field (with material damping, but without material fluctuations)
and the Green tensor appear. Without external sources, one gets
\begin{eqnarray}
    \tens{E}_{ij}( {\bf x}, {\bf x}'; \omega ) 
    & = & 
    \tens{E}^{\rm (free)}_{ij}( {\bf x}, {\bf x}'; \omega ) 
    \label{eq:LTE-el-spectrum}
    \\
    &  & {} +
    \int\limits_{V}\!{\rm d}^3r \,
    [G_{ik}({\bf x}, {\bf r}; \omega)]^*
    G_{jk}({\bf x}', {\bf r}; \omega)
    \frac{ 2 \hbar 
    {\rm Im}\,\varepsilon_0 \varepsilon( {\bf r}; \omega )
    }{ \exp[ \hbar\omega / k_{\rm B} T({\bf r}) ]
    - 1} 
    \nonumber
.
\end{eqnarray}
See Henry and Kazarinov for a similar approach~\cite{Henry96}.
The first term is nonzero for a bounded material surrounded by
a non-absorbing dielectric (like free space) and describes the photons 
incident from infinity towards the observation points 
${\bf x}, {\bf x}'$. It accounts for all of the field fluctuations 
when there is no material absorption at all. If the field in the
surrounding medium as assumed to be at zero temperature (like in the
visible frequency range), this term is zero for $\omega > 0$.
Even at finite temperature, however,
this term is typically negligible at sub-wavelength distances 
from an absorbing structure. Under these conditions, 
the second one dominates i.e., the radiation due to the polarization 
noise~(\ref{eq:local-FD-polarization}). We summarize explicit 
examples above planar substrates in the next Section.

\paragraph{Example: thermal van der Waals force.}
Consider a planar, dielectric substrate at 
temperature $T_{1}/k_{B} > 0$, with the empty half-space above  
at zero temperature. 
This describes a thermal source surrounded by a vacuum chamber 
with absorbing walls at zero temperature \cite{Henry96}.
(We neglect, of 
course, the heating of the walls due to the radiation from the 
substrate.) This model is the idealization of a situation where the 
dielectric is heated to a temperature higher than its surroundings. 

The fluctuations of the electromagnetic field
now have two contributions: a first one coming 
from the thermal currents in the substrate, and a second one coming 
from the vacuum fluctuations in the empty half-space. Written
schematically,
\begin{equation}
    \langle E_{i}({\bf r}; \omega) E_{j}^{\dag}({\bf r}'; \omega')
    \rangle = 
    2\pi\delta(\omega - \omega')
    \left\{
    W_{ij}[\omega; T_{1}; {\bf r}, {\bf r}']
    +
    V_{ij}[\omega; T = 0; {\bf r}, {\bf r}']
    \right\}
    .
    \label{eq:LTE-field}
\end{equation}
There is no cross term because the two sources are not correlated.
The spectrum $W_{ij}$ describes the radiation of the substrate
and is computed using the Lifshitz model: it is the radiation
produced by thermal current fluctuations inside the substrate.
The spectrum $V_{ij}$ describes the vacuum field in the empty half-space.
It may be calculated \`a la Lifshitz by allowing for a nonzero
imaginary part into the vacuum dielectric constant that is put to
zero in the final result \cite{Henry96,Welsch95}. Alternatively,
one can perform an explicit field quantization 
in the half-space \cite{Mandel71} and retain only the modes
incident from the vacuum side. The result is the same in both cases,
of course.

The contributions in~\eref{eq:LTE-field}
can be combined in the following way to recover the
equilibrium situation at $T_1=0$.  We recall that for $\omega < 0$, the
substrate field spectrum is proportional to $1 + \bar n( |\omega| / 
T_{1} )$, while
the vacuum spectrum is proportional to $1 + 0 = 1$.  Similarly, for
positive frequencies, we have $W_{ij}( \omega ) \propto \bar n( 
\omega / T_{1} )$, and
$V_{ij} = 0$.  Schematically, we may write
\begin{equation}
    W_{ij}[\omega; T_{1}; {\bf r}, {\bf r}']
    =
    W_{ij}[\omega; T = 0; {\bf r}, {\bf r}'] + 
    \bar n(|\omega| / T_{1}) W_{ij}[|\omega|; T = 0; {\bf r}, {\bf r}']
    \label{eq:separate-thermal}
\end{equation}
where the second term vanishes at zero temperature and is even in 
$\omega$. 
Those terms in~\eref{eq:LTE-field} that survive at 
$T_{1} = 0$ combine to give the zero temperature
fluctuation-dissipation theorem.

The radiation emitted by the substrate 
(the second term in~\eref{eq:separate-thermal})
gives the explicitly temperature-dependent radiation.
Note that this term is not the imaginary part of the Green tensor 
because we are not dealing with an equilibrium situation. Its 
behaviour as a function of frequency and the interpretation of the 
corresponding atom-surface force is given in~\cite{Henkel02a}.

\end{small}

\subsection{Example: planar surface}

In this section, we calculate the properties of the electromagnetic
field close to a planar surface. The fluctuation--dissipation theorem
is used, and a plane wave expansion of the Green tensors for the
electric and magnetic fields is worked out. Power law asymptotics
are determined. This text is adapted from the habilitation thesis
``Coherence theory of atomic de Broglie waves and electromagnetic near 
fields'' (C. Henkel, Universit\"{a}t Potsdam, April 2004).

\begin{small}

\subsubsection{Electric and magnetic Green tensors}
\label{a:Weyl}

Consider a nonmagnetic solid with (relative)
permittivity $\varepsilon$ and
permeability $\mu = 1$ that fills the half-space $z \le 0$.
We shall be interested in the fluctuation spectrum of the
electromagnetic field in the vacuum half-space, in
particular for sub-wavelength distances 
$0 < z < \lambda \equiv 2\pi c / \omega$. Up to hundreds of
nanometers from the surface, this regime is relevant even
at optical frequencies. 

For the Green tensor $\tens{G}( {\bf x}, {\bf x}'; \omega)$,
we can make the \emph{ansatz}, provided both ${\bf x}$ and 
${\bf x}'$ are located outside the solid,
\begin{equation}
\tens{G}( {\bf x}, {\bf x}'; \omega)
=
\tens{G}^{\rm (vac)}( {\bf x}, {\bf x}'; \omega)
+
\tens{G}^{\rm (refl)}( {\bf x}, {\bf x}'; \omega)
\label{eq:G-hom+refl}
\end{equation}
where $\tens{G}^{\rm (vac)}$ is the vacuum Green 
tensor. 
$\tens{G}^{\rm (refl)}$ describes the electric field reflected
from the solid and is determined from the boundary conditions for
the electric and magnetic fields at $z = 0$. 
The decomposition~(\ref{eq:G-hom+refl})
is convenient to compute the field fluctuation spectra via the
FD theorem~(\ref{eq:FD-theorem}) because it exhibits clearly the
additional contribution due to the scattering from the 
surface~\cite{Sipe84,Agarwal75a}.
This statement remains true for scatterers of arbitrary shape,
with more a complicated expression for the reflected or scattered field,
of course.

Above a planar solid, the so-called Weyl expansion or angular spectrum
representation provides a natural plane-wave basis for the incident 
and reflected fields, see the textbook by Nieto-Vesperinas~\cite{Nieto}. 
The $xy$-plane naturally plays a distinguished role here. 
Introducing two-dimensional in-plane wave vectors ${\bf Q}
= (q_{x}, \, q_{y})$, we shall 
use the notation 
\begin{equation}
{\bf q}( \pm ) =
{\bf Q} \pm {\bf n} q_z
.
\end{equation}
where ${\bf n}$ is the unit normal. 
One then has the following Fourier expansion for the reflected
Green tensor~\cite{Sipe84}
\begin{eqnarray}
    \mbox{$z, z' > 0$:}
    && 
    \tens{G}^{\rm (refl)}( {\bf x}, {\bf x}' ; \omega ) 
    = 
    \label{eq:refl-G-tensor}
    \\
    && \frac{ {\rm i} \omega^2 }
    { 2 \varepsilon_{0} c^2 }  
    \int\!\frac{ {\rm d}^2Q }{ (2\pi)^2}
    \frac{ {\rm e}^{{\rm i} {\bf q}(+) \cdot {\bf x} - 
    {\rm i} {\bf q}(-) \cdot {\bf x}' } }
    { q_{z} }
    \sum_{\mu \, =\, {\rm s,\,p}}
    r_{\mu}
    {\bf e}_{{\mu}}( + ) {\bf e}_{{\mu}}( - )
,
\nonumber
\end{eqnarray}
where the integral runs over all in-plane wave vectors ${\bf Q}$.
The wave
vectors for downward and upward waves (both propagating and
evanescent) are given by ${\bf q}( - )$ and ${\bf q}( + )$, 
respectively.
The disc $|{\bf Q}| \le \omega/c$ corresponds to propagating
waves where
\begin{equation}
q_z = \sqrt{ (\omega/c)^2 - Q^2 },
\qquad
{\rm Im}\,q_z \ge 0
\label{eq:def-qz}
\end{equation}
is real, while
$| {\bf Q}Ê| > \omega/c$ describes evanescent waves that decay or increase 
exponentially with distance. Evanescent waves are required in the
Green tensor to describe correctly the near field of a point dipole.
For the reflected field, they provide the 
dominant contribution at subwavelength distances from the solid. 
All elementary plane waves satisfy the vacuum dispersion relation
$( {\bf q}(\pm) )^2 = (\omega/c)^2$. 
Their polarization vectors are given by
\begin{eqnarray}
{\bf e}_{\rm s}( \pm ) &=& \hat{\bf Q} \times {\bf n}
,
\nonumber
\\
{\bf e}_{\rm p}( \pm ) &=& 
\frac{ {\bf q}( \pm ) \times {\bf e}_{\rm s} }{ \omega / c }
=
\frac{ \pm \hat{\bf Q} q_z - {\bf n} Q }{ \omega / c }
,
\label{eq:def-pol-vectors}
\end{eqnarray}
where $\hat{\bf Q}$ is the unit vector along ${\bf Q}$.
We have normalized the polarization vectors~(\ref{eq:def-pol-vectors})
such that ${\bf e}^2_{\mu}( \pm ) = 1$; note that
no complex conjugation is involved, although the vectors are complex 
in general. The conventional polarizations s and p are also 
called TE and TM in the literature (TE = electric field transverse 
to the plane of incidence spanned by ${\bf n}$ and $\hat{\bf Q}$).

The reflection from the solid mixes downward with upward waves
and is characterized by the Fresnel reflection 
coefficients~\cite{Jackson,BornWolf} 
\begin{eqnarray}
r_{\rm s} &=& 
\frac{ 
q_z - \sqrt{ \varepsilon(\omega) (\omega/c)^2 - Q^2 } 
}{ 
q_z + \sqrt{ \varepsilon(\omega) (\omega/c)^2 - Q^2 } 
}
\\
r_{\rm p} &=& 
\frac{ 
\sqrt{ \varepsilon(\omega) (\omega/c)^2 - Q^2 } 
- \varepsilon(\omega) q_z 
}{ 
\sqrt{ \varepsilon(\omega) (\omega/c)^2 - Q^2 } 
+ \varepsilon(\omega) q_z 
}
\label{eq:defs-refl-Green}
\end{eqnarray}
Note that the permittivity of the solid only enters via the
Fresnel coefficients. As long as the planar symmetry is not 
broken, Eq.~(\ref{eq:refl-G-tensor}) can also be used above a 
multilayer medium~\cite{Yeh,Tomas95}.
Note that sign conventions differ for the Fresnel coefficients
and polarization vectors;
only the product of $r_{\mu}$ and the polarization vectors
appearing under the sum in Eq.~(\ref{eq:refl-G-tensor})
has an unambiguous meaning.

We define the magnetic Green tensor by analogy to 
Eq.~(\ref{eq:def-Green}) as the magnetic induction field radiated by 
a point magnetic moment 
\begin{equation}
    {B}_{i}( {\bf x}; \omega ) =  
    \tens{H}_{ij}( {\bf x}, {\bf x}'; \omega ) \mu_{j}
.
\label{eq:B--Hm}
\end{equation}
From the Maxwell equations 
we find that in terms of 
its electric counterpart, the magnetic Green tensor in the vacuum
above the solid is given by the double curl
\begin{equation}
    \tens{H}_{ij}( {\bf x}, {\bf x}'; \omega ) =
    \frac{1}{\omega^2 }
    \epsilon_{ikl}\epsilon_{jmn}
    \frac{\partial}{ \partial x_{k} }
    \frac{\partial}{ \partial x'_{m} }
    \tens{G}_{ln}( {\bf x}, {\bf x}'; \omega )
    .
\label{eq:Hij-Gij}
\end{equation}
For the free space Green tensor, this leads to an expression similar 
to the electric case. 
For the reflected field, as given by the Green 
tensor~(\ref{eq:refl-G-tensor}), we observe that the curl
exchanges the polarization vectors~(\ref{eq:def-pol-vectors})
according to
\begin{equation}
    {\bf q} \times {\bf e}_{\rm s} = \frac{ \omega }{ c } {\bf e}_{\rm p},
    \qquad
    {\bf q} \times {\bf e}_{\rm p} = - \frac{ \omega }{ c } {\bf e}_{\rm s}
\label{eq:mag-pol-vecs}
\end{equation}
because ${\bf q}$, ${\bf e}_{\rm s}$, and ${\bf e}_{\rm p}$ form 
an orthogonal \emph{Dreibein}. Hence, up to a factor $1/c^2$, we 
obtain the magnetic field reflected from the solid by exchanging 
the reflection coefficients $r_{\rm s} \leftrightarrow r_{\rm p}$
in Eq.~(\ref{eq:refl-G-tensor}).

\subsubsection{Short distance expansions}

%

To illustrate the behaviour of the field at short distances,
we review here asymptotic expansions for the electric and
magnetic Green tensors. 

\paragraph{Electric field.}
As a first step, we show that in the near field, the 
reflected part~(\ref{eq:refl-G-tensor}) of the electric 
Green tensor takes a simple, electrostatic form. 
The integral over
the wave vector ${\bf Q}$ involves the factor ${\rm e}^{ {\rm i} 
q_{z} (z + z')}$ which provides a natural
cutoff for large ${\bf Q}$ as soon as $q_{z}$ becomes imaginary with 
$|q_{z}| \ge 1/(z+z') \gg \omega / c$.
Analyzing the integrand, we notice that it peaks around the 
cutoff value. We thus get the leading order asymptotics
by using an expansion for $Q, |q_{z}|$ much larger than $\omega / c$
under the integral. 

Let us first assume the more stringent condition
$Q \gg |\sqrt{\varepsilon}| \omega/c$ which corresponds
to a distance much shorter than the medium wavelength,
$z \ll \lambda / |\sqrt{\varepsilon}|$. The reflection coefficients
and polarization vectors then behave like
\begin{eqnarray}
    r_{\rm s} \,
        {\bf e}_{{\rm s}}( + ) {\bf e}_{{\rm s}}( - )
	&\approx&
	( \varepsilon - 1 )
	\frac{ \omega^2 }{ 4 Q^2 c^2 }
	( {\bf n} \times \hat{\bf Q} )_{}
	( {\bf n} \times \hat{\bf Q} )_{} 
,
\label{eq:rs-xnf}
\\
    r_{\rm p} \,
        {\bf e}_{{\rm p}}( + ) {\bf e}_{{\rm p}}( - )
	&\approx&
	\frac{\varepsilon - 1}{\varepsilon + 1}
	\frac{ Q^2 c^2 }{ \omega^2 }
	( {\bf n} - {\rm i} \hat{\bf Q} )_{}
	( {\bf n} + {\rm i} \hat{\bf Q} )_{}
,
\label{eq:rp-xnf}
\end{eqnarray}
where ${\bf n}$ is the surface normal. 
The polarizations behave very differently, the p-polarized 
part dominating for large $Q$ by a factor $( Q c / \omega )^4$.
The reflection coefficient $r_{\rm p}$ 
tends towards the electrostatic value
$r_{\rm stat} = (\varepsilon - 1)/(\varepsilon + 1)$ and becomes
independent of $Q$. 
This means that the reflection from the surface is 
nondispersive and can be modeled in terms of image 
theory~\cite{Jackson}:
the reflected field corresponds to the well-known
field of an \emph{image dipole} $\tilde{\bf d}
= (-d_{x}, - d_{y}, d_{z})r_{\rm stat}$ located at the position
${\bf X}' - z'{\bf n}$ below the surface.
Performing the integrals over ${\bf Q}$, we indeed find the
short-distance asymptotics
\begin{eqnarray}
    &&\tens{G}^{\rm (refl)}( {\bf X}' + {\bf R}, z, {\bf X}', z' ;
    \omega ) 
    \label{eq:refl-G-xnf}\\
    &&    \approx 
    \frac{\varepsilon( \omega ) - 1}{\varepsilon( \omega ) + 1}
    \frac{ 
    (R^2 + \bar z^2) \mathbbm{1} - 3 {\bf R} {\bf R} 
    + (\bar z^2 - 2 R^2) {\bf n} {\bf n}
    + 3 \bar z\,({\bf R} {\bf n} - {\bf n} {\bf R}) }{
    4\pi\varepsilon_{0} \, (R^2 + \bar z^2)^{3/2} }
,
\nonumber
\end{eqnarray}
where 
$\bar z = z+z'$ and $(R^2 + \bar z^2)^{1/2}$ is 
the distance between the observation point and the image dipole.
This expression depends
on frequency only via the electrostatic reflection coefficient.
Note also the broken isotropy of the correlation tensor where the 
coordinates parallel and perpendicular to the surface appear
in non-equivalent ways. Nevertheless, the symmetry 
relation~(\ref{eq:symmetry-Green}) 
is satisfied.


In a similar way, an asymptotic expression for distances
larger than $\lambda / |\sqrt{\varepsilon}|$ can be worked out.
This is particularly interesting above metallic surfaces 
where $\varepsilon \approx {\rm i} \sigma / \varepsilon_0
\omega$ can be very large. 
In this case, the expansion~(\ref{eq:refl-G-xnf}) is 
valid for $z \ll \delta_{\omega}$ where the skin depth
\begin{equation}
\delta_{\omega} \equiv \sqrt{ \frac{2 \varepsilon_0 c^2 
}{ \sigma \omega } }
\end{equation}
can be much smaller than the wavelength (typically,
at frequencies below the infrared). For the complementary range 
$\delta_{ \omega } \ll z$, 
one finds, repeating the analysis leading to the 
asymptotics~(\ref{eq:rs-xnf}), (\ref{eq:rp-xnf}) 
(see~\cite{Henkel99c})
\begin{equation}
    {\rm Im}\,\tens{G}( {\bf X}, z, {\bf X}, z; \omega)
    \approx 
    \frac{ \omega^2 \delta_\omega \mathbbm{1} 
    }{ 32\pi\varepsilon_{0}c^2 \, z^2 } 
    .
    \label{eq:refl-G-skin}
\end{equation}
For simplicity we give only the imaginary part of the
tensor with coinciding positions. Note the different power 
law with distance $z$ and also 
the isotropic noise strength regarding the field polarization.

To summarize, the electric field fluctuation spectrum derived from the
Green tensor~(\ref{eq:refl-G-xnf}) using the FD
theorem~(\ref{eq:FD-theorem}) is
\begin{equation}
\tens{E}( {\bf x}, {\bf x}; \omega ) =
\frac{ \hbar (\omega \delta_{\omega} / c)^2 
}{ 16\pi\varepsilon_0 \,({\rm e}^{ \beta \hbar \omega } - 1) }
\left\{
\begin{array}{ll}
\displaystyle 
\frac{ \mathbbm{1} + {\bf n} {\bf n} }{ z^3 }
& \mbox{ for $z \ll \delta_{\omega}$,}
\\[1.5ex]
\displaystyle 
\frac{ \mathbbm{1} }{ \delta_{\omega} \,z^2 }
& \mbox{ for $\delta_{\omega} \ll z \ll 
(\delta_{\omega} \lambda)^{1/2}$,}
\\[1.5ex]
\displaystyle 
\frac{ 16 \omega \,\mathbbm{1} }{ 3 c \delta_{\omega}^2 }
& \mbox{ for $(\delta_{\omega} \lambda)^{1/2} \le z$.}
\end{array}
\right.
\label{eq:nf-elect-asymptotics}
\end{equation}
It is only at distances $z \ge (\delta_{\omega} \lambda)^{1/2}$ 
that the blackbody spectrum, originating from
the free space contribution (third case in 
Eq.(\ref{eq:nf-elect-asymptotics})),
becomes the dominant contribution. Closer to the surface,
the electric field fluctuations have a noise spectrum that
can exceed the Planck formula by several orders of magnitude.
Note also that the low frequency 
limit is given by a constant spectrum $\propto k_{\rm B} T / ( \sigma
\, z^3 )$. The electric field fluctuations thus behave like white 
noise on the nanometer scale.

\paragraph{Magnetic field.}
%
Analogous calculations give for the Green tensor of the magnetic
field at distances below the skin depth, $z, z' \ll \delta_\omega$,
\begin{eqnarray}
    &&\tens{H}^{\rm (refl)}( {\bf X}' + {\bf R}, z, {\bf X}', z' ;
    \omega ) 
\label{eq:mag-Green-xnf}\\
    &&    
    \approx 
\frac{ \omega^2 \mu_0 }{ 4\pi c^2 }
\left\{
\frac{ \varepsilon( \omega ) - 1 }{ 4 \tilde r ( \tilde r + \bar z )}
\left[ 
\tilde r \,\mathbbm{1} - {\bf R} {\bf R} / ( \tilde r + \bar z )
 + \bar z \,{\bf n} {\bf n} + ({\bf R} {\bf n} - {\bf n} {\bf R})
\right]
\right.
\nonumber\\
&& \quad {} \left.
+
\frac{ \varepsilon( \omega ) - 1 }{\varepsilon( \omega ) + 1}
\frac{ 
\bar z \,\mathbbm{1} + {\bf R} {\bf R} / ( \tilde r + \bar z )
 - \bar z \,{\bf n} {\bf n}
}
{ \tilde r ( \tilde r + \bar z ) }
\right\}
,
    \nonumber
\end{eqnarray}
where $\tilde{r} = ( {R}^2 + \bar z^2 )^{1/2}$.
From the fluctuation-dissipation theorem, we get the 
fluctuation spectrum of the magnetic field. We give here also the
regimes of larger distances
\begin{equation}
\tens{B}( {\bf x}, {\bf x}; \omega ) =
\frac{ \hbar \mu_{0} }{ 8\pi\,\delta_{\omega}^2 
\,({\rm e}^{ \beta \hbar \omega } - 1) }
\left\{
\begin{array}{ll}
\displaystyle 
\frac{ \mathbbm{1} + {\bf n} {\bf n} }{ z }
& \mbox{ for $z \ll \delta_{\omega}$,}
\\[1.5ex]
\displaystyle 
\frac{ \delta_{\omega}^3 (\mathbbm{1} + {\bf n} {\bf n}) }{ 3 z^4 }
& \mbox{ for $\delta_{\omega} \ll z \ll 
(\delta_{\omega} \lambda^3)^{1/4}$,}
\\[1.5ex]
\displaystyle 
\frac{ 8 \delta_{\omega}^2 \omega^3 \,\mathbbm{1} }{ 3 c^3 }
& \mbox{ for $(\delta_{\omega} \lambda^3)^{1/4} \le z$.}
\end{array}
\right.
\label{eq:nf-mag-asymptotics}
\end{equation}
Note the different exponents for the power laws with distance
compared to the electric field,
and the larger cross-over distance to the blackbody radiation
spectrum. The low-frequency limit of the magnetic noise spectrum
is $\propto \mu_{0}^2 k_{\rm B} T \sigma / z$, it is 
frequency-independent as well.  Similar expressions
have been derived in~\cite{Sidles00,Henkel99c,Varpula84}.

\subsubsection{Electromagnetic energy densities}
\label{s:nf-noise-discussion}

Some of the results summarized above have been discussed by Joulain,
Carminati, Mulet, and Greffet in a recent paper on the definition and
measurement of the \ldos{} close to planar surfaces~\cite{Joulain03}.
These authors
analyze the spectrum of the electric and magnetic energy densities
$u_{\rm e}( z ; \omega )$, $u_{\rm m}( z ; \omega )$
as a function of distance and point out that the 
definition 
           of the \ldos{}
should be taken with care given the
non-equivalent role played by the electric and magnetic fields in the 
near field. To illustrate this, we plot in 
Figure~\ref{fig:em-nf-noise} the ratio $u_{\rm m}( z ; \omega ) 
/ u_{\rm e}( z ; \omega )$. Notice the strong dominance of the magnetic 
energy throughout the near field range up to $z \sim \lambda$ for
a metallic surface.
The asymptotic formulas~(\ref{eq:nf-elect-asymptotics}), 
(\ref{eq:nf-mag-asymptotics}) provide good agreement with a numerical 
calculation based on the exact plane wave expansion~(\ref{eq:refl-G-tensor}) 
for the electric Green functions and its magnetic equivalent.

\begin{figure}
\centerline{\resizebox{!}{48mm}{%
\includegraphics*{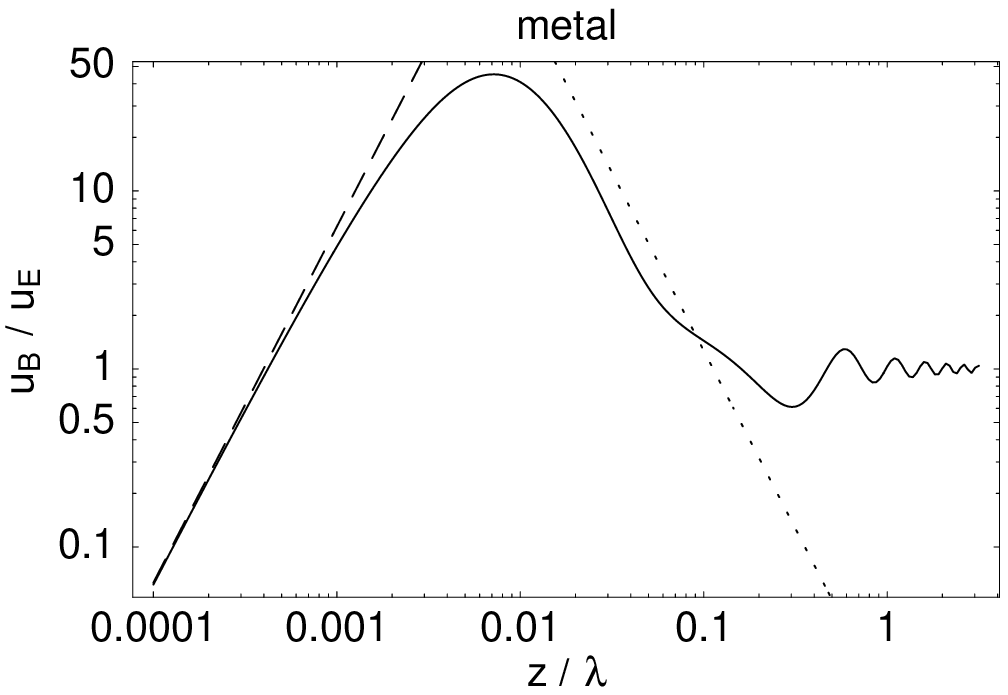}}
\resizebox{!}{48mm}{%
\includegraphics*{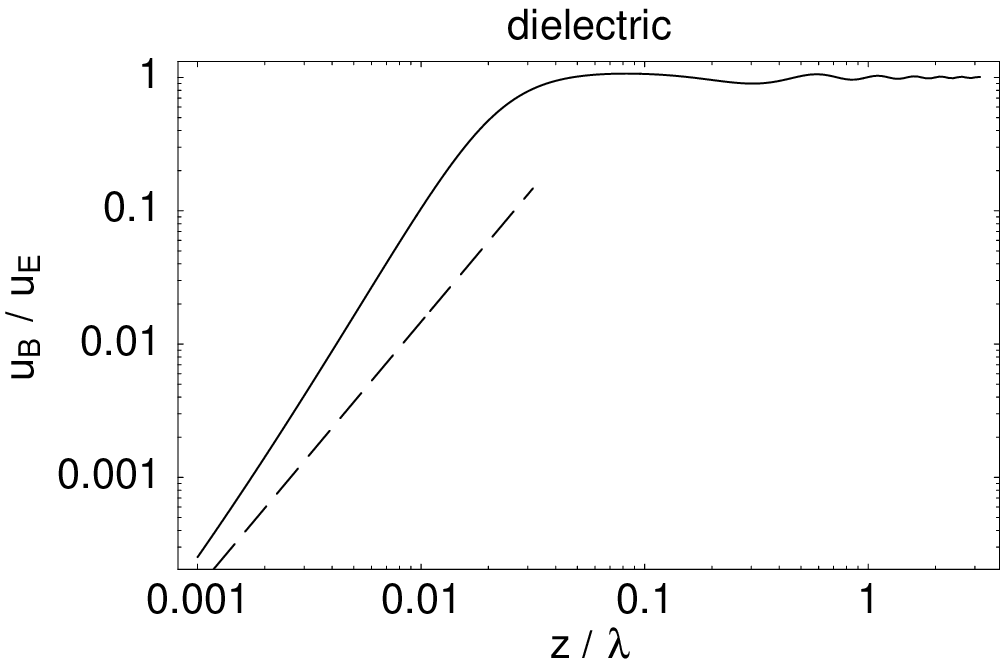}}}
\caption[]{\footnotesize 
Ratio of magnetic to electric energy density
vs.\ distance from a medium-filled half-space, 
normalized to the wavelength $\lambda$. 
Left panel: metal with $\varepsilon = 1 + 800\,{\rm i}$
(skin depth $\delta = 0.008\,\lambda$).
Right panel: dielectric ($\varepsilon = 2.3 + 0.1\,{\rm i}$,
$\delta > \lambda$).
The dashed and dotted lines 
correspond to the asymptotic 
expansions~(\ref{eq:nf-elect-asymptotics}), 
(\ref{eq:nf-mag-asymptotics}).}
\label{fig:em-nf-noise}
\end{figure}

For a dielectric surface with an essentially real permittivity,
the near field energy is dominantly electric, as shown in
Fig.~\ref{fig:em-nf-noise} (right). This behavior is not covered by the 
asymptotics~(\ref{eq:nf-elect-asymptotics}), 
(\ref{eq:nf-mag-asymptotics}) because the assumption 
${|\varepsilon|} \gg 1$ breaks down, but can be found 
from~(\ref{eq:refl-G-xnf}) and its magnetic counterpart. 
At distances comparable to the wavelength, oscillations appear for both
metallic and dielectric surfaces, that correspond to the standing waves
formed by the partial reflection of the field.  In the far field regime
$z \gg \lambda$, the symmetry between electric and magnetic energy is
restored, as expected from the Planck formula.

\end{small}

\subsection{Discussion of the van der Waals force}

\subsubsection{Short distance limit}

The dynamic polarizability for an atomic ground state has been given
in~(\ref{eq:alpha-ground-state}). A transition frequency $\omega_{eg}$ 
provides a natural scale for the $\omega$-integral of the van der 
Waals force. As a first estimate, we can assume that frequencies 
around the strongest electric dipole transition, $\omega \sim 
\omega_{eg}$ contribute most strongly to the integral. This allows to 
define the short distance limit: the distance is small compared to 
the corresponding (reduced) wavelength, $z \ll \lambdabar_{eg} \equiv
c / \omega_{eg}$. Note that the same physical limit is achieved by 
neglecting retardation, $c \to \infty$, or focussing on low 
frequencies, $\omega \sim \omega_{eg} \to 0$. 

Now, the low frequency (small distance) 
limit of the Green tensor is simply given by electrostatics. For a 
planar surface, the reflected field can be constructed in terms of an 
image dipole located below the surface at a symmetric point. From 
this construction, one finds a Green tensor
\begin{equation}
    {\cal G}_{ij}^{({\rm sc})}( {\bf r}, {\bf r}; \omega ) =
    \frac{1 }{4\pi\varepsilon_{0}}
    \frac{ \varepsilon( \omega ) - 1 }{ \varepsilon( \omega ) + 1 }
    \frac{ \delta_{ij} + \delta_{iz} \delta_{jz} }{ (2z)^3 }
    ,
\end{equation}
as also outlined around Eq.(\ref{eq:refl-G-xnf}). The van der Waals 
interaction energy is now given by
\begin{eqnarray}
E_{\rm vdW}( z ) &\approx&
- \frac{ 1 }{ 2 \pi\varepsilon_0 (2z)^3}
\sum_{e} \left(
| \langle g | {\bf d} | e \rangle |^2 
+
| \langle g | {d}_z | e \rangle |^2 
\right)
\nonumber
\\
&& {}\times \,{\rm Im}
\int\limits_0^\infty\!\frac{ {\rm d}\omega }{ 2 \pi }
\frac{ \varepsilon( \omega ) - 1 }{ \varepsilon( \omega ) + 1 }
\frac{ \omega_{eg} }{ \omega_{eg}^2 - \omega^2 - 0 {\rm i} \omega}
.
\label{eq:vdw-shift-near-field}
\end{eqnarray}
This expression is used to interpret high-resolution atomic spectra 
near surfaces, see the lecture by D. Bloch. Particularly interesting 
features occur when the atomic resonance, $\omega \sim \omega_{eg}$, 
gets close to a ``surface resonance'', $\varepsilon( \omega ) 
\sim - 1$. If a model for the dielectric function valid at complex 
frequencies is known, one can also evaluate the integral in~(\ref{eq:vdw-shift-near-field})
along the imaginary axis:
\begin{equation}
    \int\limits_0^\infty\!\frac{ {\rm d}\xi }{ 2 \pi }
    \frac{ \varepsilon( {\rm i}\,\xi ) - 1 }{ \varepsilon( {\rm i}\,\xi ) + 1 }
    \frac{ \omega_{eg} }{ \omega_{eg}^2 + \xi^2 }
    .
    \label{eq:vdw-int-imag-freq}
\end{equation}
All quantities here are real and positive so that no real or imaginary 
part is needed any more.%
\footnote{A response function at $\omega = {\rm i}\,\xi$ corresponds 
to the response to a real-valued excitation that is exponentially 
switched on, $\propto {\rm e}^{ \xi t}$. Since the physical response 
is a real (hermitean) variable, the response function is real as well.}



\subsubsection{Large distance limit}

At large distances, we have to use the full Green tensor. It can be 
written in the form
\begin{equation}
{\cal G}^{({\rm sc})}_{ij}( {\bf r}, {\bf r}; \omega ) 
=
\frac{ {\rm i} }{ 
8\pi \varepsilon_0 }
\int\limits_{0}^{\infty}\!K \,{\rm d}K 
\frac{ {\rm e}^{ 2 {\rm i} \gamma z } }{ \gamma }
\left\{
\frac{ \omega^2 }{ c^2 }
\Delta_{ij}\,
r_{\rm s}( \varepsilon )
+
\left(
K^2 \delta_{iz} \delta_{jz}
- \gamma^2 
\Delta_{ij}
\right)
r_{\rm p}( \varepsilon )
\right\}
\label{eq:full-el-Green}
\end{equation}
where $\Delta_{ij} = {\rm diag}( 1, 1, 0 )$ in a frame where the 
$z$-axis points normal to the surface. The distance occurs in the 
exponential
\begin{equation}
\exp( 2 {\rm i} \gamma z )
=
\exp( - 2 z \sqrt{ K^2 - \omega^2 / c^2 } )
.
\end{equation}
One shifts the $\omega$-integration onto the imaginary axis and 
observes that in the limit $z \to \infty$, only small values of $K$ 
and $\omega = {\rm i}\,\xi$ contribute. We can thus expand the 
integrand for small $K$ and $\xi$ and evaluate the integral. 
We find to lowest order the static atomic polarizability 
$\alpha_{ij}^{({\rm stat})}$.
The combination $\kappa = \sqrt{ K^2 + \xi^2 / c^2 }$ is a good integration 
variable, and the integral gives the famous $1/z^4$ power law of the 
``Casimir--Polder'' interaction. The 
remaining integral can be written in terms of an angle $\alpha$ 
(with $\xi = \kappa \,\cos\alpha$), and we get
\begin{eqnarray}
    V_{\rm CP}( z ) &=& - \frac{ 3 \hbar c \alpha_{ij}^{({\rm stat})} }{
    8 \pi^2 \varepsilon_{0} (2z)^4 } \times {}
    \label{eq:CP-shift}
    \\
    &&{}\times
    \int\limits_0^{\pi/2}\!{\rm d}\alpha\,
    \sin\alpha 
    \Big\{
    - \cos^2\alpha \,
    \Delta_{ij}\,
    r_{\rm s}( \varepsilon_{{\rm stat}} )
    +
    \left(
    2 \sin^2\alpha \,
    \delta_{iz} \delta_{jz}
    + 
    \Delta_{ij}
    \right)
    r_{\rm p}( \varepsilon_{{\rm stat}} )
    \Big\}.
    \nonumber
\end{eqnarray}
The reflection coefficients are evaluated with the static dielectric 
constant, but depend in general on $\alpha$, for example:
\begin{equation}
    r_{\rm s}( \varepsilon_{{\rm stat}} ) =
    \frac{ 1 - \sqrt{ \varepsilon_{{\rm stat}} \cos^2 \alpha + \sin^2 
    \alpha }  }{ 1 + \sqrt{ \varepsilon_{{\rm stat}} \cos^2 \alpha + \sin^2 
    \alpha } }
    \label{eq:rs-CP-limit}
\end{equation}
For a perfect conductor and also for metals with a nonzero \textsc{dc} 
conductivity, $\varepsilon_{{\rm stat}} = {\rm i}\,\infty$, and
therefore $r_{\rm s} \equiv -1$ and $r_{\rm p} \equiv +1$. In that 
case, the $\alpha$-integral gives $\frac43 \delta_{ij}$.


\section{Transitions}
\label{s:atom-chips}

In this second part of the lecture, we discuss transitions between 
different quantum states of a trapped atom that are induced by 
electromagnetic fluctuations. For atoms trapped close to surface, the 
relevant quantum states are: Zeeman or hyperfine sublevels
that are trapped in potentials of different steepness (or not trapped 
at all), and center-of-mass eigenstates in the trapping potential. In 
the first case, transitions lead to trap loss or to the decoherence of 
Zeeman or hyperfine superposition states. In the second case, field 
fluctuations feed energy into the center-of-mass motion and give 
heating.

The transition rates can be computed, as a first approximation, with 
second-order perturbation theory, using Fermi's Golden Rule. We shall 
find out that the rates are indeed small so that higher order 
corrections are usually not needed. We show in particular that the 
transition rate is a meausre of the cross spectral density of the 
field fluctuations. Let us look first at the transition between different
magnetic sublevels. The following section is adapted from the paper 
``Fundamental limits for coherent manipulation on atom chips'' by C.\ 
Henkel,  P.\ Kr\"uger, R.\ Folman, and J.\ Schmiedmayer [Appl. 
Phys. B {\bf 76} (2003) 173].

\begin{small}

\subsection{Noise spectrum and spin flip rate}

The coupling of the atomic magnetic moment to fluctuating magnetic
fields gives rise to both spin flips and changes in the center-of-mass
motion (scattering). The rate of these processes is given by
Fermi's Golden Rule. We recall here that it can be conveniently
expressed in terms of the noise spectrum of the magnetic field
fluctuations.
(See \cite{Sipe84} for a similar approach and Chap.\ IV
of \cite{CDG2} for the derivation of a full master equation.)

If we write $|{\rm i} \rangle$ and $|{\rm f}\rangle$ for the atomic
states before and after the transition, the transition rate is
\begin{equation}
\Gamma_{\rm i \to f} =
\frac{ 2 \pi }{ \hbar }
\sum_{\rm F,I}
p( {\rm I} )
\left|
\langle {\rm F, f} |
H_{\rm int}
| {\rm I, i} \rangle
\right|^2
\delta(
E_{\rm F}
+
E_{\rm f}
-
E_{\rm I}
-
E_{\rm i}
)
,
\label{eq:Gamma-if-Fermi}
\end{equation}
where $|{\rm I} \rangle$ and $|{\rm F}\rangle$ are initial and final
states for the field, the summation being an average over the initial
field states (with probabilities $p({\rm I})$) and a trace over the final
field states.  The interaction Hamiltonian is given by
$H_{\rm int} = - \mbox{\boldmath$\mu$}\cdot{\bf B}( {\bf x} )$.

Consider first the rate for spin flips. Since only a
subset of magnetic sublevels $|m_{\rm i}\rangle$ are weak field
seekers, spin flips $|m_{\rm i}\rangle \to |m_{\rm f}\rangle$
are responsible for trap loss. The magnetic field is
evaluated at the position ${\bf r}$ of the trap center. (An
average over the atomic position distribution would be more
accurate.)
We write the $\delta$-function for energy conservation as a time
integral
over ${\rm e}^{{\rm i}(
E_{\rm I}
-
E_{\rm F}
-
\hbar \omega_{\rm f\-i})
t / \hbar}$
where
$\hbar\omega_{\rm f\-i} = E_{\rm f} - E_{\rm i}$.
The exponential ${\rm e}^{ {\rm i} ( E_{\rm I} - E_{\rm F})
t / \hbar}$
can be removed by introducing
the field operators in the Heisenberg picture
\begin{equation}
{\bf B}( {\bf r}, t ) =
{\rm e}^{ {\rm i} H_0 t / \hbar }
{\bf B}( {\bf r} )
{\rm e}^{ -{\rm i} H_0 t / \hbar }
\end{equation}
(here, $H_0$ is the free field Hamiltonian)
and taking matrix elements of this operator between the initial
and final field states. This gives
\begin{equation}
{\rm e}^{ {\rm i} ( E_{\rm I} - E_{\rm F}) t / \hbar }
\langle {\rm I} |
{\bf B}( {\bf r} )
| {\rm F} \rangle
=
\langle {\rm I} |
{\bf B}( {\bf r}, t )
| {\rm F} \rangle
.
\end{equation}
The sum over the final states $| {\rm F}\rangle$ now reduces
to a completeness relation and we get ($\alpha,\beta$ denote
field components)
\begin{eqnarray}
&&
2\pi\hbar
\sum_{\rm F,I}
p( {\rm I} )
\langle {\rm I} |
{B}_\alpha( {\bf r} )
| {\rm F} \rangle
\langle {\rm F} |
{B}_\beta( {\bf r} )
| {\rm I} \rangle
\delta(
E_{\rm F}
-
E_{\rm I}
- \hbar \omega
)
\nonumber
\\
&& =
\int\limits_{-\infty}^{\infty}
\!{\rm d}t \,
{\rm e}^{ {\rm i}\omega t }
\sum_{\rm I}
p( {\rm I} )
\langle {\rm I} |
{B}_\alpha( {\bf r}, t )
{B}_\beta( {\bf r}, 0 )
| {\rm I} \rangle
\nonumber
\\
&& =
{\cal B}_{\alpha\beta}( {\bf r}, {\bf r}; \omega )
.
\label{eq:def-noise-spectrum}
\end{eqnarray}
In the last line, we have defined the magnetic noise spectrum
which is the Fourier transform of the
field's autocorrelation function.
The rate for spin flips can now be written as
\begin{equation}
\Gamma_{\rm i \to f} =
\frac{ 1 }{ \hbar^2 }
\sum_{\alpha, \beta = x,y,z}
\langle m_{\rm i} | \mu_\alpha | m_{\rm f} \rangle
\langle m_{\rm f} | \mu_\beta | m_{\rm i} \rangle
{\cal B}_{\alpha\beta}( {\bf r}, {\bf r}, -\omega_{\rm f\-i} )
.
\label{eq:rate-spectrum}
\end{equation}
Since $m_{\rm f} \ne m_{\rm i}$, the matrix elements of
$\mu_{\alpha}$ are only nonzero for directions perpendicular to the
magnetic field at the trap center. We also recover the selection rule
$m_{\rm f} - m_{\rm i} = \pm 1$ so that the relevant transition
frequency is the Larmor frequency $|\omega_{\rm f\-i}| = \omega_{\rm L}$.
The spin flip rate gives the order of
magnitude of trap loss even if more than one weak-field seeking
Zeeman states, $m_{\rm i} = +2, +1$, say, are trapped
(possible with many of the alkali atoms).
This is because the
matrix elements between adjacent sublevels do not significantly differ
in magnitude so that the atoms reach the non-trapped sublevel
$m_{\rm f} = 0$ after a time $ \sim 2/\Gamma_{+2 \to +1}$.

We finally note that as long as the behaviour of the
`environment' (the field) is ignored in the description
of the atom's dynamics, the noise spectrum is the only quantity
needed to characterize the environment.
It is also an experimentally
measurable quantity: for example, the rms magnetic noise
$\langle B_{x}^2( {\bf r} ) \rangle^{1/2}$
measured by a spectrum analyzer in a given frequency band
$\Delta \omega / 2\pi$ around $\omega$
is $( 2 {\cal B}_{xx}( {\bf r}, {\bf r}; \omega ) \, \Delta \omega / 2\pi )^{1/2}$,
the factor $2$ accounting for the sum over positive and negative
frequencies. The atomic spin flip rate may be regarded as
an alternative way to measure the noise spectrum. In order
of magnitude,
the magnetic moment is comparable to the Bohr magneton,
$\mu_{\rm B}$ 
($\mu_{\rm B} / 2\pi\hbar = 1.4\,$MHz/G), and we get
\begin{equation}
\Gamma_{\rm i\to f}( {\bf r} ) \sim
0.01\,{\rm s}^{-1}
( \mu / \mu_{\rm B} )^2
\frac{ {\cal B}_{\alpha\beta}( {\bf r}, {\bf r}; \omega_{\rm L} ) } {
{\rm pT}^2 / {\rm Hz} }
.
\end{equation}
Note that current SQUID magnetometers are able to detect magnetic field
noise even on the $10\,{\rm fT}/\sqrt{ {\rm Hz} }$ scale
\cite{Varpula84}.

\end{small}

\subsection{Magnetic near field noise}

Eq.(\ref{eq:rate-spectrum}) clearly shows that we need the spectrum 
of the magnetic field fluctuations for an estimate of the spin flip 
transition rate. The fluctuation--dissipation theorem provides the 
spectrum in terms of the Green function 
${\cal H}_{ij}( {\bf x}, {\bf x}'; \omega )$
for the magnetic field,
\begin{equation}
    {\cal B}_{ij}( {\bf x}, {\bf x}'; \omega ) = \frac{ 2\hbar }{ {\rm 
    e}^{\hbar\omega/T} - 1}
    {\rm Im}\,{\cal H}_{ij}( {\bf x}, {\bf x}'; \omega )
    .
    \label{eq:FD-magnetic}
\end{equation}
The regime typical for trapped atom experiments is sub-wavelength 
distances, where the magnetic Green tensor (remember: the field 
radiated by a point magnetic moment) is dominated by the part 
reflected or scattered from the surface. From the short-distance
expansion~(\ref{eq:mag-Green-xnf}), we find two regimes, depending on 
the relatize magnitude of the distance compared to the skin depth 
inside the surface material.

\paragraph{Very short distance, normal conductor.}
The skin depth $\delta_{\omega}$ is large compared to the distance.
As a scaling law for a normal metal,
\begin{equation}
    \delta_{\omega} = 160\,\mu{\rm m} \left( 
    \frac{ \varrho / \varrho_{\rm Cu} }{ \omega / 2\pi \, {\rm MHz} }
    \right)^{1/2}
    ,
    \label{eq:skin-depth-scaling}
\end{equation}
where $\varrho$ is the specific resistance. The scaling of the 
spin flip rate is then
\begin{equation}
    \Gamma_{\rm i\to f} \propto
    \frac{ T }{ \varrho \,z}
    ,
    \label{eq:flip-rate-scaling-xnf}
\end{equation}
This estimate applies to a half-space. For absolute numbers, see the 
lecture by J. Schmiedmayer. Note in particular that the flip rate is 
independent of the Larmor frequency. This is due to a conspiracy 
between the ``density of states'' of the magnetic near field 
($\propto \omega$) and the 
Bose-Einstein occupation number ($\propto 1/\omega$).

The estimate~(\ref{eq:flip-rate-scaling-xnf}) can be extended to other
geometries using the following argument. In this ``very short distance''
regime, the distance $z$ gives the depth of the material that contributes 
significantly to the magnetic noise outside the surface. This has 
been checked by an explicit calculation in~\cite{Varpula84,Henkel01a}. 
Any structure thicker than $z$ thus essentially behaves like a half-space.
One gets a reduction of the magnetic noise spectrum for thinner 
films or wires, for example. Each dimension along which the metallic 
material becomes ``finite'' (characteristic length $a$ like thickness 
or wire radius) gives an additional factor $a/z$ for the noise 
spectrum and hence for the spin flip\ rate. It is likely that this 
explains the cross-over seen in the lifetime data vs.\ distance
of J.\ Schmiedmayer's group.

If one works in the short-distance regime, the optimal material choice 
is an insulator (large resistance $\varrho$). 
For strong magnetic fields,
permanently magnetized structures can be used. Alternatively, 
materials may exist whose conductivity is reasonably large in the
\textsc{dc} range and sufficiently small in the kHz to MHz range
that applies for the relevant transitions. 

\paragraph{Intermediate distance, very good conductor.}
At larger distances or equivalently, smaller resistance, the skin depth 
ultimately becomes smaller than the distance. In this regime, a 
different power law is obtained. One gets a steeper scaling with 
distance,
\begin{equation}
    \Gamma_{\rm i\to f} \propto
    \frac{ T \,\varrho^{1/2} }{ \omega_{\rm L}^{1/2} \,z^4}
    ,
    \label{eq:flip-rate-scaling-skin}
\end{equation}
which is now also frequency dependent. This regime should also apply 
to superconducting substrates. Note that in the relevant frequency 
range, even a superconductor shows some finite resistance due to the 
normal electron fraction not bound into Cooper pairs.

In this regime, one can say that qualitatively, only the surface of 
the material produces magnetic field fluctuations. The power laws for
different geometries therefore do not probably show 
the cross-over seen at short distance. The problem is currently under 
study. Estimates for magnetic noise above superconducting substrates 
are given in the review by J.\ A.\ Sidles~\cite{Sidles00}.



%

\subsection{Other rates}

The rate estimates for other processs like heating (excitation of higher 
trap eigenstates) and decoherence (decay of a superposition of 
magnetic sublevels into a statistical mixture) are discussed in the 
review paper by R.\ Folman and colleagues~\cite{Folman02} and in the
review by C.\ Henkel and the J.\ Schmiedmayer group~\cite{Henkel03a}.
The most robust choice for superposition states are magnetic sublevels 
that have the same magnetic moment. They are shifted in the same way 
by magnetic fluctuations so that their relative phase is preserved. 
Such states exists in alkali atoms, they have the same orientation of 
the electron spin and opposite nuclear spins. The difference in 
magnetic moment is only provided by the nuclear magneton which is 
much smaller than the Bohr magneton, by roughly the mass ratio between 
the nucleon and the electron.

\section*{Conclusion}

Atoms are a sensitive probe of electromagnetic fields close to surfaces. 
Their energy level shifts depend on the reflection or scattering of 
the field from the surface, integrated over the entire frequency range. 
At short distances, the range around atomic transition frequencies 
gives the dominant contribution. This can lead, for a ``hot'' surface,
to interesting non-equilibrium forces. 

Transitions between different ``internal'' atomic states or the 
heating in a trap close to a surface gives directly access to the 
field's cross spectral density at the transition frequency, provided 
perturbation theory holds. In magnetic traps that confine only a subset 
of Zeeman sublevels, the magnetic spectrum is the dominant driving 
agent for transitions and leads to trap loss, heating, and decoherence. 
The corresponding rates
show different characteristic scalings with distance and substrate 
geometry, depending on the size of the skin depth inside the material.


\begin{footnotesize}
    
\bibliography{\bibpath journals,\bibpath biblioac,\bibpath bibliodh,%
\bibpath biblioio,\bibpath bibliopz,%
extrabib}
\bibliographystyle{\bstpath prsty}

\end{footnotesize}

\end{document}